\documentclass{emulateapj}

\usepackage{graphicx}
\usepackage{rotating}
\usepackage{longtable}
\usepackage{threeparttablex}
\usepackage{verbatim}
\usepackage{appendix}
\usepackage{subfigure}

\newcommand{\lapprox }{{\lower0.8ex\hbox{$\buildrel <\over\sim$}}}
\newcommand{\gapprox }{{\lower0.8ex\hbox{$\buildrel >\over\sim$}}}
\newcommand{\Msun}{\ensuremath{M_{\odot}}}
\newcommand{\Lsun}{\ensuremath{L_{\odot}}}

\newcommand{\Teff}{\ensuremath{T_{\mathrm{eff}}}}

\newcommand{\Macc}{\ensuremath{\dot M_{\mathrm{acc}}}}
\newcommand{\Lacc}{\ensuremath{L_{\mathrm{acc}}}}
\newcommand{\brg}{Br$\gamma$}
\newcommand{\pab}{Pa$\beta$}
\newcommand{\pag}{Pa$\gamma$}

\newcommand{\AH}{\ensuremath{A_{\mathrm{H}}}}

\slugcomment{Accepted to PASP: 18 September 2012}
\shorttitle{Oph SpeX Variability}
\shortauthors{Faesi et al.}

\begin{document}

\title{Potential Drivers of Mid-infrared Variability in Young Stars:\\Testing physical models with multi-epoch near-infrared spectra of YSOs in $\rho$~Oph}

\author{Christopher M. Faesi\altaffilmark{1,2}, Kevin R. Covey\altaffilmark{3,4,5}, Robert Gutermuth\altaffilmark{6}, Maria Morales-Calder\'{o}n\altaffilmark{7}, John Stauffer\altaffilmark{7}, Peter Plavchan\altaffilmark{8}, Luisa Rebull\altaffilmark{7}, Inseok Song\altaffilmark{9}, James P. Lloyd\altaffilmark{3} }

\altaffiltext{1}{Harvard University, Department of Astronomy, 60 Garden Street, Cambridge, MA 02138}
\altaffiltext{2}{NSF Graduate Research Fellow}
\altaffiltext{3}{Cornell University, Department of Astronomy, 226 Space Sciences Building, Ithaca, NY 14853}
\altaffiltext{4}{Hubble Fellow}
\altaffiltext{5}{Visiting Astronomer at the Infrared Telescope Facility, which is operated by the University of Hawaii under Cooperative Agreement no. NNX-08AE38A with the National Aeronautics and Space Administration, Science Mission Directorate, Planetary Astronomy Program}
\altaffiltext{6}{Department of Astronomy, University of Massachusetts, Amherst, MA 01003, USA}
\altaffiltext{7}{Spitzer Science Center, 1200 E California Blvd., Pasadena, CA 91106, USA }
\altaffiltext{8}{NASA Exoplanet Science Institute, California Institute of Technology, Pasadena, CA 91125, USA}
\altaffiltext{9}{University of Georgia, Department of Physics and Astronomy, Athens, GA 30602}

\begin{abstract}

Recent studies have identified several young stellar objects (YSOs) which exhibit significant mid-infrared (mid-IR) variability. A wide range of physical mechanisms may be responsible for these variations, including changes in a YSO's accretion rate or in the extinction or emission from the inner disk. We have obtained and analyzed multi-epoch near-infrared (NIR) spectra for five actively accreting YSOs in the $\rho$~Oph star-forming region along with contemporaneous mid-IR light curves obtained as part of the YSOVAR \textit{Spitzer}/IRAC survey. Four of the five YSOs exhibit mid-IR light curves with modest ($\sim 0.2$--$0.4$~mag) but statistically significant variations over our 40-day observation window. Measuring the strengths of prominent photospheric absorption lines and accretion sensitive \ion{H}{1} and \ion{He}{1} lines in each NIR spectrum, we derive estimates of each YSO's spectral type, effective temperature ({\Teff}), and $H$~band extinction ($A_H$), and analyze the time evolution of their NIR veiling ($r_H$ and $r_K$) and mass accretion rates ({\Macc}). Defining a YSO's evolutionary stage such that heavily veiled, high accretion rate objects are less evolved than those with lower levels of veiling and ongoing accretion, we infer that GY~314 is the most evolved YSO in our sample, with GY~308 and GY~292 at progressively earlier evolutionary stages. Leveraging our multi-epoch, multi-wavelength dataset, we detect significant variations in mass accretion rates over timescales of days to weeks, but find that extinction levels in these YSOs remain relatively constant. We find no correlation between these YSO mid-IR light curves and time-resolved veiling or mass accretion rates, such that we are unable to link their mid-IR variability with physical processes localized near the inner edge of the circumstellar disk or within regions which are directly responsive to mass accretion. We do find, however, that redshifted \ion{He}{1}~$\lambda 10830$ emission, where present in our spectra, shows both quantitative and qualitative temporal correlations with accretion-sensitive \ion{H}{1} emission lines. Blueshifted \ion{He}{1} absorption, on the other hand, does not demonstrate a similar correlation, although the time-averaged strength of this blueshifted absorption is correlated with the time-averaged accretion rate in our sample of YSOs.

\end{abstract}


\section{Introduction}
\label{sec:intro}

Young stars have long been known to demonstrate significant photometric variability (e.g., \citealp{Joy45}). Variability at optical and near-infrared (NIR) wavelengths is often attributed to phenomena on or near the photosphere, such as hot and cold spots produced by accretion and stellar magnetic fields \citep[e.g.,][]{Herbst94,JB95,CHS01,AJB01,Bouvier07}. Recent observations of young stellar objects (YSOs) have also identified significant variability in the mid-infrared \citep[mid-IR; e.g.,][]{Liu96,BRM05,MoralesCalderon09,MoralesCalderon10,Muzerolle09,Flaherty11,Flaherty12}, where a YSO's spectral energy distribution (SED) is thought to be dominated by emission from a circumstellar disk. Dramatic changes have been seen over timescales as short as days, including variations in the shape of the infrared disk continuum \citep{Juhasz07,MoralesCalderon09,Muzerolle09,Flaherty11} and in the silicate feature, which is expected to trace the dust in the inner disk \citep{Barsony97,Sitko08,BLS09,Skemer10}.

These recent detections of YSO mid-IR variability have been driven by the sensitivity and angular resolution provided by the \textit{Spitzer} Space Telescope \citep{Werner04}. While \textit{Spitzer} is now in the post-cryogenic phase of its mission, the 3.6 and 4.5 $\mu$m channels of its InfraRed Array Camera \citep[IRAC;][]{Fazio04} continue to provide high sensitivity to emission from the inner regions of primordial circumstellar disks. The YSOVAR (Young Stellar Object VARiability) \textit{Spitzer} Exploration Science Program (PI:~Stauffer) is the first large-scale systematic survey of YSO variability in the mid-IR. The recently published YSOVAR light curves of $\sim2000$ objects in the Orion Nebula Cluster demonstrate a wide range of morphologies; many YSOs exhibit aperiodic variability in the mid-IR, such that evolutionary state appears correlated with both the likelihood and amplitude of variability. Class I objects are most likely to be variable, and exhibit the largest amplitudes, with Class II   YSOs (a.k.a. classical T~Tauri stars, or CTTSs) and Class III YSOs exhibiting progressively fewer, and weaker, variations \citep{MoralesCalderon11}.

Several scenarios have been suggested to explain YSO mid-IR variability, such as variable mass accretion \citep{Vorobyov09}, changes in extinction \citep{Herbst94}, large structural changes in the inner disk \citep{Sitko08}, changes in the height of the disk's inner wall \citep{Espaillat11}, stellar occultation by density enhancements in the inner disk \citep{MoralesCalderon11}, and disk turbulence \citep{TCS10}. Discriminating between these models, however, requires that the mid-IR light curves be supplemented with contemporaneous multi-wavelength observations. If accretion rate variations drive the observed mid-IR variability, for example, the strength of accretion-sensitive spectral lines should show changes commensurate with the objects' light curves; indeed, \cite{LR99} (hereafter LR99) observed simultaneous variability in {\brg} emission and $K$~band veiling\footnote[1]{Veiling in the infrared is a measure of excess emission (generally interpreted as being due to the presence of a circumstellar disk), and is typically defined as $r_\lambda \equiv F_{\mathrm{excess},\lambda}/F_{*,\lambda}$, where $F_{\mathrm{excess},\lambda}$ and $F_{*,\lambda}$ are the excess continuum flux (i.e. disk emission) and stellar flux at wavelength $\lambda$, respectively.} in two YSOs in $\rho$~Oph, hinting that accretion activity and disk emission may be connected in young stars. Extinction and/or emission from localized structures in the disk may also contribute to mid-IR variability; simultaneous mid-IR photometry and NIR spectroscopy, such as the observations we present in this study, allow these scenarios to be tested.

$\rho$~Oph is one of the nearest ($d \sim 120\pm5$~pc; \citealp{Loinard08}) star-forming regions, and has been extensively catalogued in the optical and NIR~(e.g. LR99; \citealp{Bontemps01,Wilking01}; see also \citealp{WGA08} and references therein). YSOs in $\rho$~Oph are notably variable at NIR wavelengths, with many Oph YSOs exhibiting aperiodic variability in the NIR on timescales of months to years (e.g. \citealp{Plavchan08}; \citealp{Scholz12}; \citealp{Parks_prep}); variability at longer wavelengths and on shorter timescales, however, is less well documented. In this work, we present a coordinated, multi-wavelength study of several active YSOs in $\rho$~Oph to explore physical mechanisms that may drive their mid-IR variability. In \S \ref{sec:observations}, we describe our photometric and spectroscopic observations and present the resultant light curves and spectra for the YSOs in our sample. In \S \ref{sec:analysis} we infer spectral types and extinction, report time series measurements of each YSO's veiling, extinction, and accretion rate, and analyze NIR/mid-IR SEDs to study disk emission. Finally, we discuss the implications of these results for understanding YSO variability, accretion, and evolution in \S \ref{sec:discussion}.

\section{Observations}
\label{sec:observations}

\subsection{Mid-IR Photometry}
\label{sec:mid-IRphot}

\begin{deluxetable*}{l l | c c | r  r  r}
\tablecolumns{7}
\tablewidth{0pt}
\tablecaption{Basic Data for Target YSOs in $\rho$~Oph}
\tablehead{
\colhead{}		& \colhead{}		& \colhead{RA}		& \colhead{Dec}	&
\colhead{$J$}			& \colhead{$H$}		& \colhead{$K_S$} \\
\colhead{Source ID}		& \colhead{other IDs}	& \colhead{(J2000)}	& \colhead{(J2000)} 	&
\multicolumn{3}{c}{(2MASS mag.)}
}
\startdata
GY 292 & ISO 155 & 16:27:33.11 & -24:41:15.3 & 11.32 & 9.13 & 7.81 \\
GY 308 & IRS 49, ISO 163 & 16:27:38.32 & -24:36:58.6 & 11.38 & 9.43 & 8.27 \\
GY 314 & ISO 166 & 16:27:39.43 & -24:39:15.5 & 10.75 & 9.21 & 8.46 \\
YLW 15a & IRS 43, ISO 141, GY 265& 16:27:26.94 & -24:40:50.8 & 18.53 & 13.52 & 9.75 \\
YLW 16b & GY 274, ISO 145 & 16:27:29.41 & -24:39:17.0 & 16.72 & 12.64 & 9.84 \\
\enddata
\label{tab:basicdata}
\end{deluxetable*}

\textit{Spitzer}/IRAC light curves were obtained for YSOs in the $\rho$~Oph star forming region as part of the YSOVAR GO6 Exploration Science program. IRAC $3.6$ and $4.5~\mu$m observations were obtained for three non-contiguous sub-fields in the $\rho$~Oph cluster between 11 April and 17 May 2010 (MJD 55298--55334); each $5.2' \times 5.2'$ dual-band sub-field was centered on an overdensity of previously known cluster members. A single observing sequence consisted of 8 visits, logarithmically spaced over 3.5 days to sample a variety of timescales and to avoid period aliasing in characterizing variability; this 3.5 day sequence was repeated 12 times to obtain monitoring data spanning $\rho$~Oph's full 40 day \textit{Spitzer} visibility window. The observations were taken in High Dynamic Range mode, with a short (0.6 s) and long (10.4 s) exposure taken at each position within a five-point gaussian dither pattern.

\begin{figure*}
\subfigure{\includegraphics[width=0.5\linewidth]{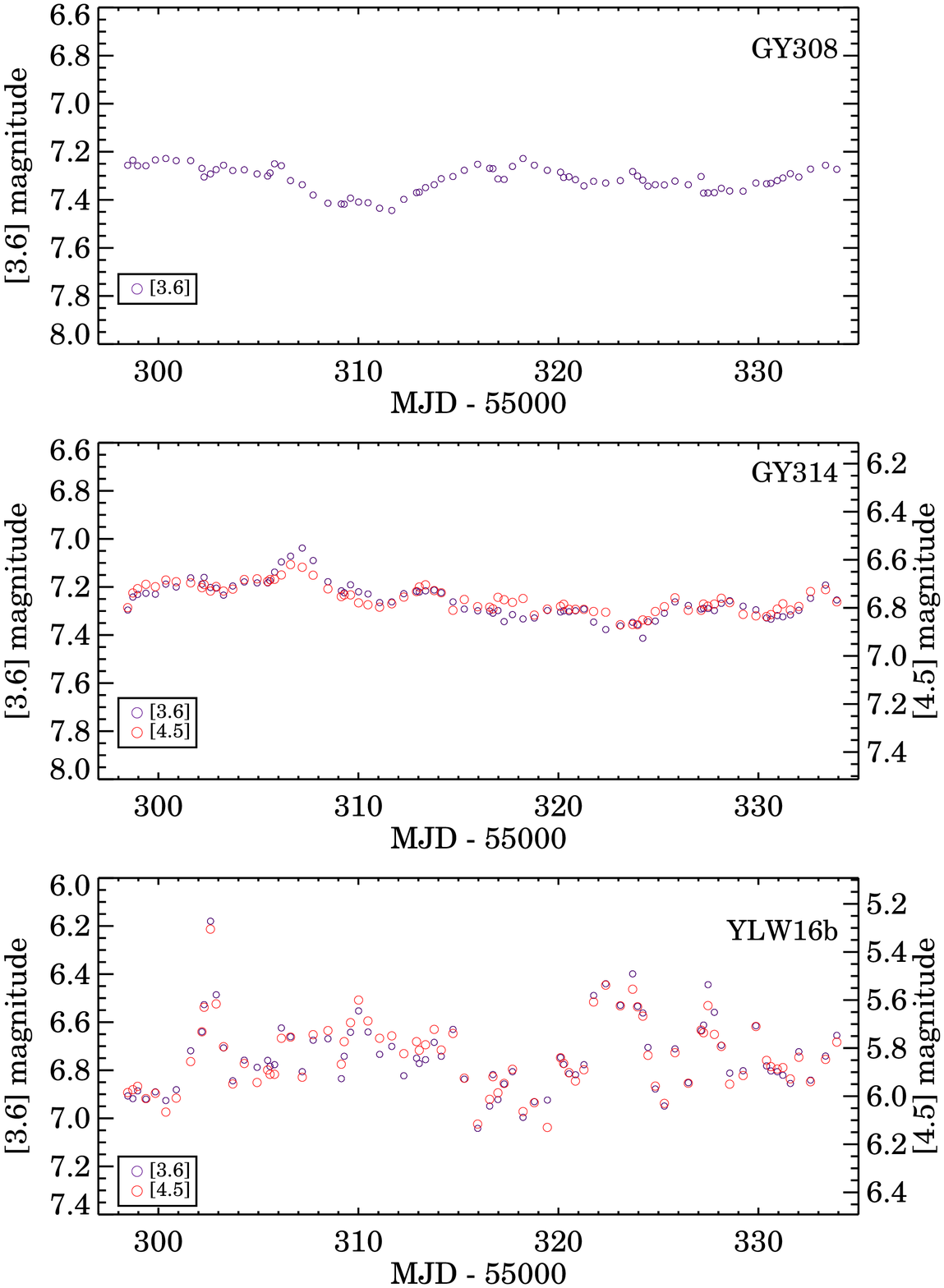}}
\subfigure{\includegraphics[width=0.5\linewidth]{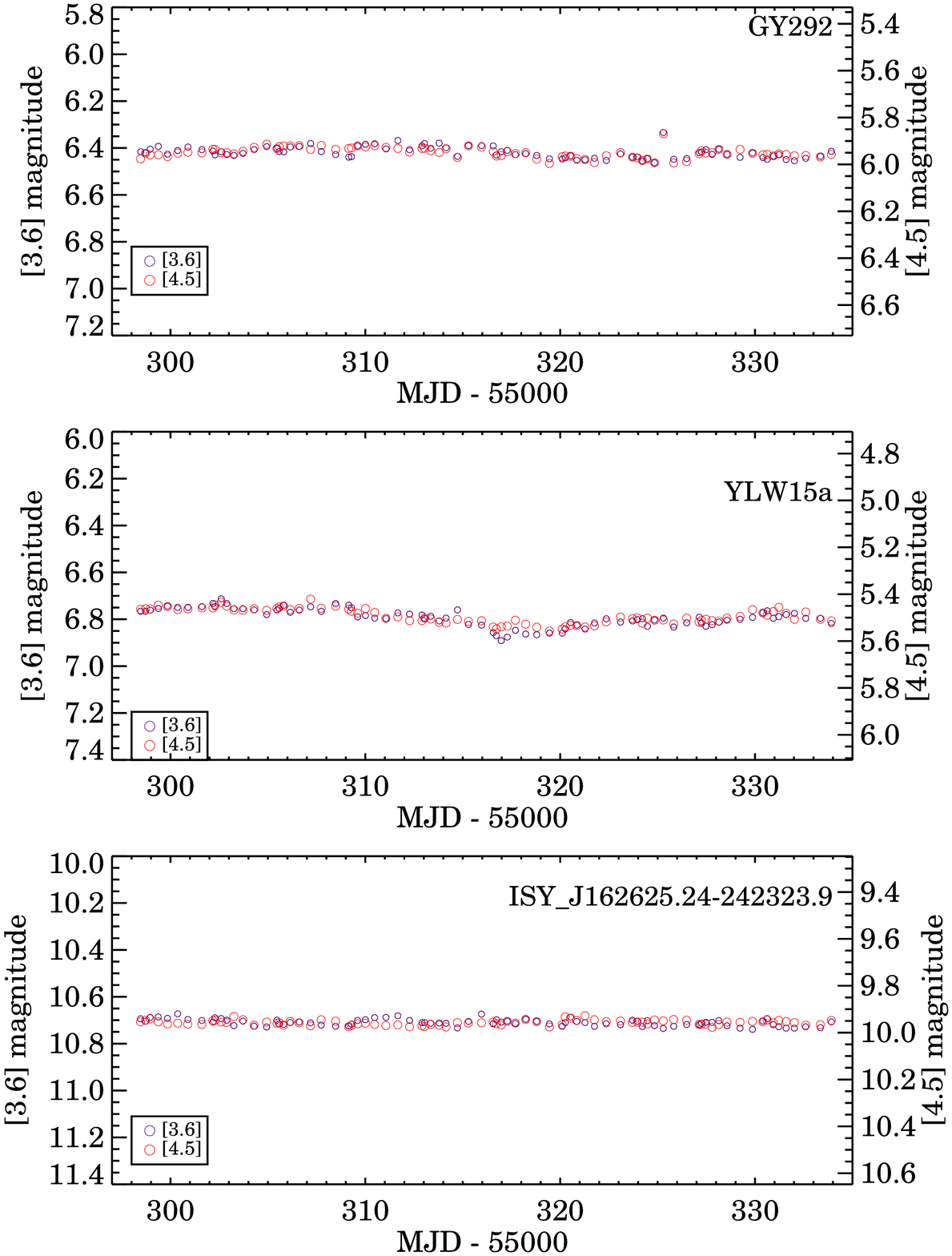}}
\caption{\small{IRAC [3.6] (purple circles, left y-axes) and [4.5] (red circles, right y-axes) light curves for the 5 YSOs analyzed here, covering the 40-day \textit{Spitzer} observation window from April 11 to May 17, 2010 (MJD 55298--55334). The left and right y-axes are shifted such that the median magnitudes in the two IRAC bands coincide. Typical photometric errors are on the order of 0.01 to 0.02 magnitudes, smaller than the plotted symbols in the figure. The three YSOs in the left panel show modest ($\sim$0.2--0.4 mag), but significant, mid-IR variations, while GY 292 and YLW 15a (right panels) are only marginally variable (see also Figure \ref{fig:iracvar}). The light curve of ISY\_J162625.24-242323.9, a nonvariable star in the YSOVAR $\rho$ Oph fields which is notably fainter than any of our science targets, is also shown to demonstrate the stability of our calibrated YSOVAR data.}}
\label{fig:lcs}
\end{figure*}

We reduced the IRAC images with the custom YSOVAR pipeline, a modified version of the algorithm originally developed by \cite{Gutermuth09} to process IRAC data for a large sample of embedded clusters. The version of the YSOVAR pipeline used to generate the light curves presented here takes as inputs the Basic Calibrated Data (BCD) images released by the Spitzer Science Center, mosaics the individual images at each epoch, and then performs source detection and aperture photometry on the combined cluster mosaics. The sets of long and short exposures at each epoch are separately mosaicked and then combined; however, for bright sources with pixels above the nominal saturation limit in any long exposure BCD frames, only the short exposure mosaics are used for photometry. Sources are then loaded into the YSOVAR database\footnote[2]{Data for the six objects analyzed in this paper are publicly available at \url{http://ysovar.ipac.caltech.edu/}}, where light curves can be constructed for individual objects by merging all detections with a 1$\arcsec$ search radius, allowing for $\sim$single-pixel centroid variations (IRAC pixels are 1.2$\arcsec$ in size). The resulting IRAC light curves for our spectroscopic targets are shown in Figure~\ref{fig:lcs}. Typical photometric uncertainties for our IRAC measurements are 0.01 to 0.02 magnitudes -- smaller than the plotted symbols in Figure~\ref{fig:lcs}. To further demonstrate the stability of these observations, we include in the figure the light curve of ISY\_J162625.24-242323.9, a much fainter, non-variable object within the $\rho$ Oph YSOVAR fields.

\subsection{NIR Spectra}
\label{sec:NIRspectra}

We selected targets for spectroscopic monitoring during the YSOVAR $\rho$~Oph campaign from the catalog of YSOs identified in previous IRAC observations of the region \citep{Gutermuth09}, which identifies 46 YSOs in the three YSOVAR/IRAC fields-of-view. Limiting the sample to YSOs with $J < 11.5$ to ensure signal-to-noise (S/N) $> 50$ across the full $JHK$ spectrum in a typical 5 minute exposure, as required to provide high confidence detections of relatively weak photospheric absorption features and measure veiling from 1--2.5~$\mu$m, further narrrows the candidates for spectroscopic monitoring to five objects. Of these, only the Class~II YSOs GY~292 and GY~314 have shown evidence of active accretion (i.e., a positive detection of {\brg} or {\pab} in spectroscopic observations by LR99 and/or \citealp{NTR06}); furthermore, these two YSOs also demonstrated the potential for accretion variability as suggested by two epochs of data from LR99. In addition, a third Class~II YSO, GY~308, also met the S/N cut and showed evidence of both accretion activity and variability. While GY~308 did not fall directly within one of the dual-band YSOVAR fields, it was nonetheless observed (though only in the $3.6~\mu$m channel) due to \textit{Spitzer's} simultaneous, offset dual-band setup. We thus adopted GY~308 as a third primary target, as well as adding two Class~I YSOs, YLW~15a and YLW~16b, from the dual-band fields as secondary targets that were observed when time permitted. Basic properties of our five YSOs are presented in Table~\ref{tab:basicdata}.

\begin{sidewaysfigure*}
\vspace{-100 mm}
\subfigure{\includegraphics[width=0.5\linewidth]{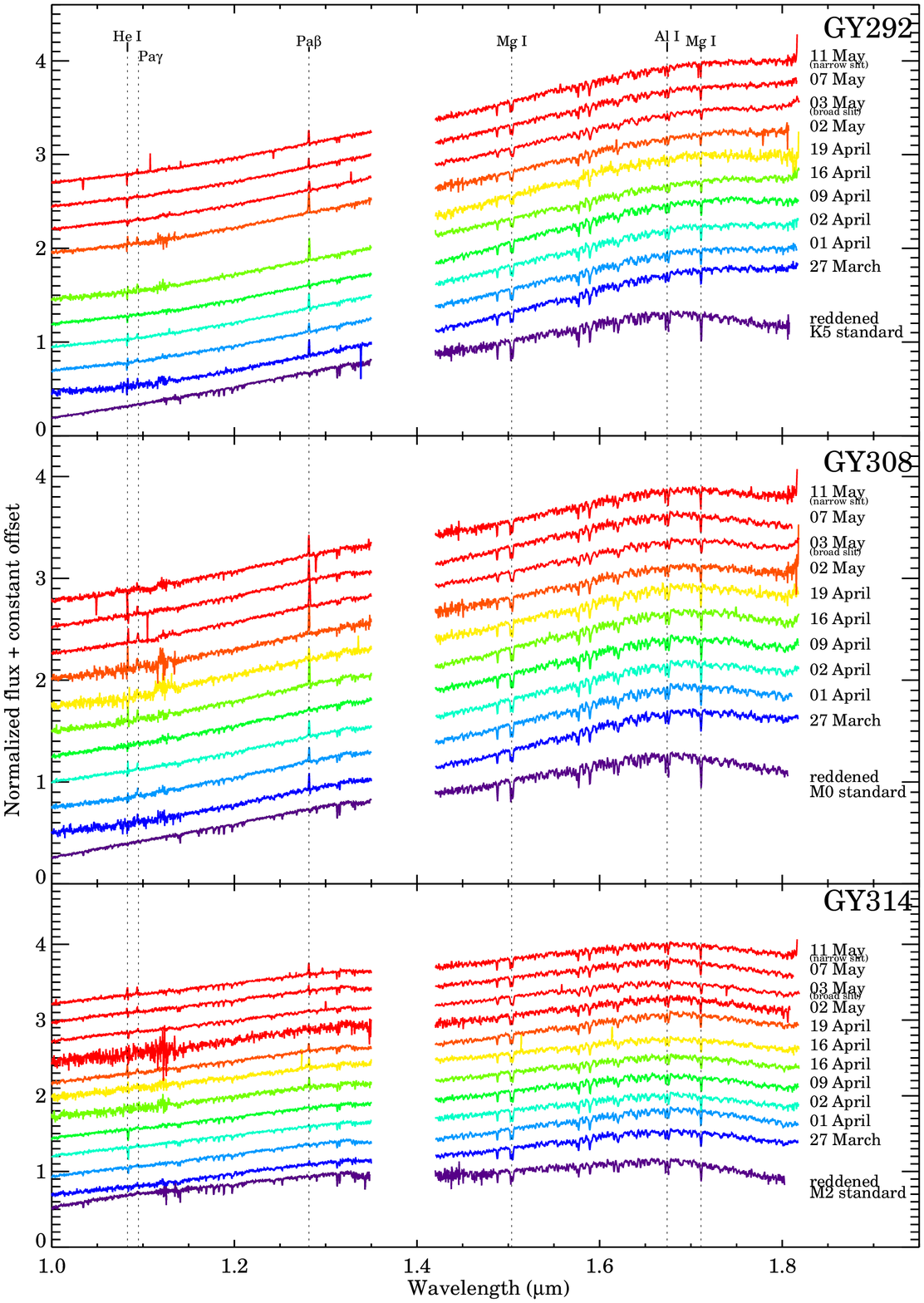}}
\subfigure{\includegraphics[width=0.5\linewidth]{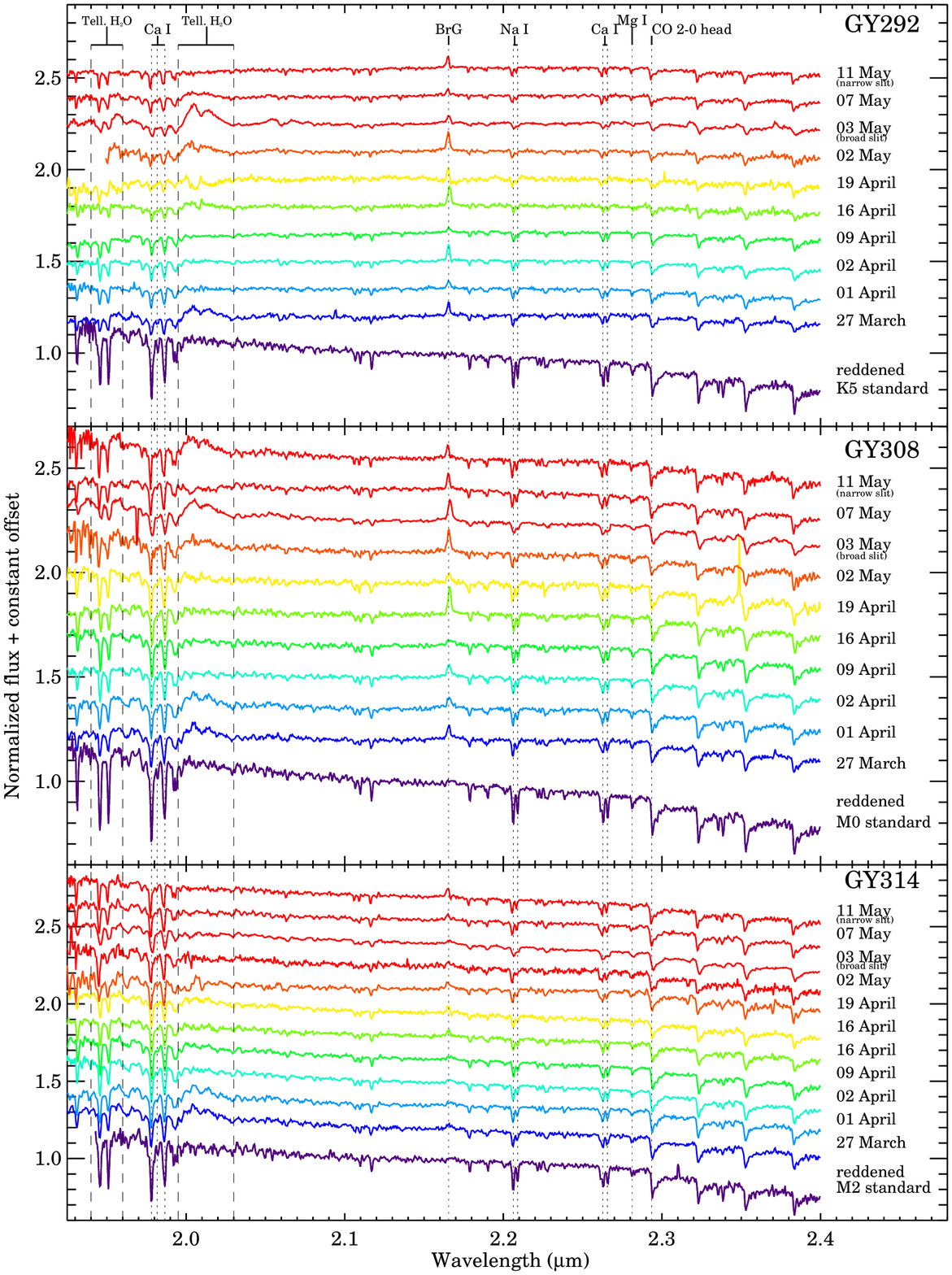}}
\caption{\small{$J$ and $H$~band (left) and $K$~band (right) spectra of GY 292, 308, and 314, with an artificially reddened spectral standard shown at the bottom of each panel for comparison. Spectra are normalized to unity at 1.5 and 2.1 {\micron} for the $J/H$ and $K$~bands, respectively. Vertical lines indicate the rest wavelengths of key features from which we infer spectral types, veiling, and mass accretion rates. Absorption lines appear weaker in the target spectra than in the standards, revealing the presence of substantial veiling flux in the target spectra; the difference in slope between the standard and target spectra is primarily due to the fact that the level of veiling is wavelength-dependent (see also Figure~\ref{fig:allveil}). Temperature-sensitive photospheric absorption lines show little evidence for variability between epochs; accretion-sensitive \ion{H}{1} emission lines, by contrast, do appear to change significantly from epoch to epoch.}}
\label{fig:spectra}
\end{sidewaysfigure*}

We used the SpeX spectrograph \citep{Rayner03} at NASA's 3.0 meter Infrared Telescope Facility (IRTF) to observe our targets on 10 nights between 27 March and 11 May 2010 (MJD 55283--55328). We present a log of our observations in Table~\ref{tab:obslog}; typical seeing was $\sim$1\arcsec, but conditions were rarely photometric, with cloud cover ranging from clear to heavily overcast. We conducted our observations with SpeX's SXD mode, providing nearly contiguous spectral coverage from 0.8 to 2.5 microns. We typically observed with a 0.5{\arcsec} slit, providing a nominal spectral resolution of R$\sim$1200 at the center of each order. On two nights of particularly good and bad seeing, we instead utilized the 0.3{\arcsec} and 0.7{\arcsec} slits, providing nominal spectral resolutions of R$\sim$2000 and $\sim$850, respectively.

\begin{deluxetable*}{l | l | r | r | c | c | c | c | c}
\tablewidth{0pt}
\tablecaption{IRTF/SpeX Observing Log and spectrally derived colors}
\tablehead{
\colhead{}		& \colhead{Observation}	\\
\colhead{}			& \colhead{Date}		& \colhead{exp.}		& \colhead{SXD}	&
\colhead{$J$}		& \colhead{$H$}		& \colhead{$K$}	
 \\
\colhead{Star}			& \colhead{(UT, 2010)}	& \colhead{time (s)}	& \colhead{slit ($\arcsec$)} &
\colhead{S/N\tablenotemark{a}}	& \colhead{S/N \tablenotemark{a}}	&
\colhead{S/N\tablenotemark{a}}	& \colhead{$J-H$\tablenotemark{b}}	&
\colhead{$H-K_s$\tablenotemark{b}}
}
\startdata
GY 292	& March 27	& 360 	& 0.5 & 46			& 168		& 295	 & 2.23 & 1.26 \\
		& April 01		& 360 	& 0.5 & 111		& 312		& 384	 & 2.19 & 1.20 \\
		& April 02		& 720 	& 0.5 & 162		& 452		& 603	 & 2.20 & 1.20 \\
		& April 09		& 720 	& 0.5 & 115		& 344		& 429	 & 2.23 & 1.26 \\
		& April 16		& 840 	& 0.5 & 60			& 172		& 209	 & 2.16 & 1.24 \\
		& April 19		& 1440 	& 0.5 & \nodata		& 83			& 155	 & 1.93 & 1.26 \\
		& May 02 		& 480 	& 0.5 & 75			& 261		& 382	 & 2.16 & 1.24 \\
		& May 03 		& 480 	& 0.7 & 132		& 375		& 512	 & 2.17 & 1.28 \\
		& May 07 		& 360 	& 0.5 & 133		& 356		& 449	 & 2.18 & 1.27 \\
		& May 11 		& 720 	& 0.3 & 132		& 334		& 391	 & 2.18 & 1.27 \\ \hline
GY 308	& March 27 	& 360	& 0.5 & 52			& 165		& 233	 & 2.04 & 1.07 \\
		& April 01 		& 360 	& 0.5 & 83			& 219		& 277	 & 1.99 & 1.03 \\
		& April 02 		& 720 	& 0.5 & 148		& 366		& 446	 & 1.98 & 1.02 \\
		& April 09 		& 720 	& 0.5 & 86			& 249		& 314	 & 1.97 & 1.00 \\
		& April 16 		& 1080 	& 0.5 & 70			& 183		& 216	 & 1.98 & 1.05 \\
		& April 19 		& 1080 	& 0.5 & 40			& 127		& 169	 & 1.97 & 1.01 \\
		& May 02 		& 840 	& 0.5 & 49			& 154		& 200	 & 2.00 & 0.91 \\
		& May 03 		& 480 	& 0.7 & 123		& 312		& 413	 & 1.91 & 1.01 \\
		& May 07 		& 360 	& 0.5 & 116		& 286		& 355	 & 1.91 & 1.00 \\
		& May 11 		& 360 	& 0.3 & 83			& 215		& 267	 & 1.91 & 1.00 \\ \hline
GY 314	& March 27	& 360 	& 0.5 & 82			& 184		& 224	 & 1.63 & 0.78 \\
		& April 01 		& 720 	& 0.5 & 166		& 310		& 324	 & 1.64 & 0.79 \\
		& April 02 		& 720 	& 0.5 & 242		& 456		& 522	 & 1.60 & 0.76 \\
		& April 09 		& 720 	& 0.5 & 170		& 333		& 375	 & 1.62 & 0.77 \\
		& April 16 		& 1260 	& 0.5 & 116		& 280		& 367	 & 1.58 & 0.80 \\
		& April 19 		& 1320	& 0.5 & 124		& 227		& 227	 & 1.66 & 0.89 \\
		& May 02 		& 960 	& 0.5 & 29			& 90			& 130	 & 1.60 & 0.78 \\
		& May 03 		& 480 	& 0.7 & 177		& 345		& 411	 & 1.57 & 0.77 \\
		& May 07 		& 360 	& 0.5 & 173		& 323		& 365	 & 1.56 & 0.79 \\
		& May 11 		& 720 	& 0.3 & 179		& 314		& 325	 & 1.59 & 0.81 \\ \hline
YLW 15a 	& March 27 	& 960	& 0.5 & \nodata		& 5			& 158	& 3.55 & 3.72 \\*
		& April 01 		& 1440	& 0.5 & \nodata		& 5			& 165	& 5.77 & 3.74 \\*
		& April 02 		& 1320	& 0.5 & \nodata		& 11			& 276	& 5.58 & 3.81\\*
		& April 09 		& 1440	& 0.5 & \nodata		& 9			& 240	& 4.74 & 3.86\\*
		& May 07 		& 600 	& 0.5 & \nodata		& 9			& 214	& 2.95 & 3.75 \\
YLW 16b	& March 27 	& 960 	& 0.5 & \nodata		& 87			& 311 	& 2.45 & 2.44 \\
		& April 01		& 1440 	& 0.5 & \nodata		& 71			& 361	& 2.54 & 2.59 \\
		& April 02 		& 960 	& 0.5 & 7			& 108		& 424	& 3.74 & 2.57 \\
		& May 07 		& 720 	& 0.5 & 13			& 131		& 362	& 3.49 & 2.30 \\
		\enddata
\tablenotetext{a}{Signal-to-noise ratios (S/N) as reported by SpeXtool.}
\tablenotetext{b}{NIR colors were computed by convolving our SpeX spectra with standard 2MASS passbands.}
\label{tab:obslog}
\end{deluxetable*}

All data were reduced using SpeX's dedicated IDL reduction package, SpeXtool \citep{CVR04}. Telluric absorption was removed using observations of A0V stars at airmasses similar to those of the target ($\Delta (\sec~z) \lesssim0.1$). Hydrogen features were removed from the A0V spectra by dividing by a model of Vega, and the target's true spectral slope and absolute flux densities were recovered by multiplying the target spectrum by a $\sim$10,000 K blackbody spectrum scaled to match that of the telluric standard \citep{VCR03}. Table \ref{tab:obslog} lists the S/N levels reported by SpeXtool for the $J$, $H$ and $K$ bands of each source's reduced SXD spectrum. We determined relative flux calibration uncertainties in these spectra by comparing independent reductions of a single raw spectrum calibrated with different telluric (and thus flux) standards. When scaled to the continuum, pairs of independent reductions agree to within 1--2\% in each band, and to within $\sim 5\%$ across the entire $JHK$ spectrum.

Our normalized SpeX NIR spectra for GY 292, 308, and 314 are shown in Figure~\ref{fig:spectra}, along with an artificially reddened spectral standard of the appropriate spectral type for each YSO (see \S \ref{sec:spt} for our spectral type estimates). The difference in continuum slope between the reddened standard and object spectra is due to veiling flux, which increases towards longer wavelengths (see Figure~\ref{fig:allveil}), in the target spectra. None of the YSO targets displayed a significant change in spectral slope over the course of the monitoring, suggesting that each YSO's veiling remained relatively constant over the course of this campaign (see \S \ref{sec:VeilVar} for a quantitative measurement of veiling and its temporal and wavelength dependence).

The secondary targets (YLW~15a and 16b) possess sufficiently high extinctions that obtaining $J$ and $H$~band spectra at useful S/N was prohibitively expensive. Moreover, we were only able to obtain $K$~band spectra for these YSOs on five and four occasions, respectively, with only one epoch for each secondary target lying within the temporal coverage of the IRAC monitoring campaign. As a result, these NIR spectra provide no information concerning the nature of the mid-IR variability exhibited by both sources; for completeness we therefore present $K$~band spectra for YLW~15a and YLW~16b in Figure~\ref{fig:ylws} of the appendix, and in the spectral archive associated with this paper, but we do not analyze these spectra in any detail.

\section{Analysis}
\label{sec:analysis}

\subsection{Mid-infrared Variability}
\label{sec:midIRvar}

\begin{figure}
\includegraphics[angle=90,width=\linewidth]{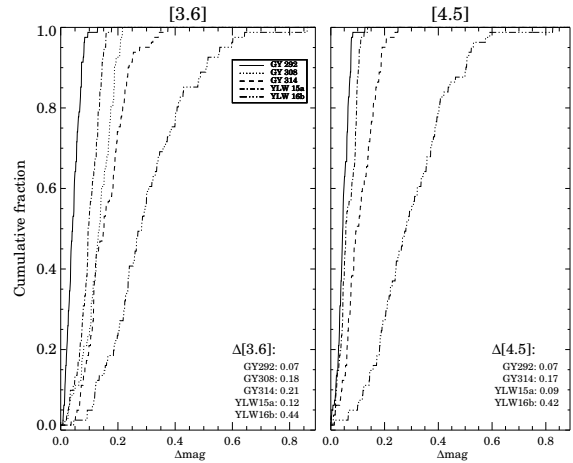}
\caption{\small{Amplitude of IRAC variability: cumulative fraction of epochs as a function of $\Delta$mag, where $\Delta$mag at a given epoch is defined as the difference between a YSO's magnitude at that epoch and at the epoch of minimum brightness. YLW 16b exhibits the largest amplitude mid-IR variability ($\gapprox 0.4$~mag), while GY 292 is only marginally variable ($\lapprox 0.1$~mag); the other YSOs demonstrate more modest, but still significant, mid-IR variations of $0.1$--$0.3$ mag. These amplitudes (which are listed on the figure as $\Delta[3.6]$ and $\Delta[4.5]$) were computed after discarding the upper and lower 5\% of the data.}}
\label{fig:iracvar}
\end{figure}

To quantify the amplitude of mid-IR variability in each source over our observational baseline, we present in Figure~\ref{fig:iracvar} cumulative histograms of the IRAC magnitudes in each light curve. Stable sources have steep slopes in this figure, and variable sources have shallower slopes, as their light curves span a larger range of magnitudes. We define the variability amplitude indices $\Delta [3.6]$ and $\Delta [4.5]$ as the range in magnitudes spanned between the 5\% and 95\% levels in the cumulative histogram for the appropriate IRAC band, i.e. the variability amplitude after discarding the upper and lower 5\% of the data. Excluding the extrema from this index guards against overestimating the variability of stable sources due to individual deviant data points. We present $\Delta[3.6]$ and $\Delta[4.5]$ for each of our five targets in Table~\ref{tab:mastertable}.

YLW 16b demonstrates the largest mid-IR variability ($\Delta[3.6]=0.44$~mag) of the five YSOs in our sample, while GY 292 shows the smallest ($\Delta[3.6]=0.07$~mag, although GY 292 appears to exhibit a short-duration flare event near MJD 55325). The mid-IR variability detected from YLW 16b, GY 314, GY 308, and YLW 15a represent $5\sigma$ excursions, with all but YLW 15a exhibiting $\Delta[3.6] \gtrsim 0.2$~mag. GY 292 is not entirely quiescent, however, as evidenced by comparing its light curve with the photometric stability demonstrated by the significantly fainter star from the same YSOVAR fields, IS\_J162625.24-242323.9, which has $\Delta[3.6] = 0.05$~mag (see Figures $\ref{fig:lcs}$ and $\ref{fig:iracvar}$). Although this small data set precludes a more quantitative statistical analysis, these results are consistent with the statistics inferred by \cite{MoralesCalderon11} from the much larger Orion YSOVAR dataset. They found that the majority of Class I and II YSOs are variable, with a large fraction having amplitudes greater than $\sim 0.2$~mag. In addition, the light curves of the most variable objects in our sample, GY~314 and YLW~16b, demonstrate strong morphological similarities to those of the irregular variables from that study.

Four of our five targets (all but GY~308) have both [3.6] and [4.5] photometry; for these YSOs, we examined the changes in the mid-IR color $[3.6] - [4.5]$ over time for evidence of a color term, which would suggest variability in the line-of-sight extinction during our monitoring program. We define $\delta_{12}$ as the observed change in $[3.6] - [4.5]$ from the median color, then compare the time evolution of this quantity with the predictions of two models: (1) the case of ``gray'' variability, i.e. $\delta_{12} \equiv 0$, and (2) the extinction law of \cite{Indebetouw05}, which predicts $\delta_{12}= (1-A_2/A_1) \delta_1$, where $A_2/A_1=0.77 \pm 0.16$ is the differential extinction in the two IRAC bands derived from Table~1 of that work, and $\delta_1$ is the change in [3.6] from the median. Computing the reduced $\chi^2$ goodness-of-fit statistic for the data as compared with each model, we find the following: for the two YSOs with more modest mid-IR variability (GY~292 and YLW~15a), both models are compatible with the data ($\chi^2 < 2$); this simply reflects the fact that the uncertainties in $\delta_{12}$ are similar in scale to the level of variability. Conversely, neither model is particularly consistent with the observed color changes ($\chi^2 \ge 6$) for YLW~16b, which shows stochastic, measurable changes in $\delta_{12}$ and no evidence for a color term. Finally, for GY~314, the extinction model is more consistent with the data ($\chi^2_{\mathrm{ext}} = 1.4$) than the gray model ($\chi^2_{\mathrm{gray}} = 5.0$), but the trend appears to be leveraged by a small subset of the data, with the vast majority actually clustered near $(\delta_1,\delta_{12})=(0,0)$. Given that $\chi^2_{\mathrm{ext}} > 1$, the relatively high uncertainty on the extinction model slope, and the fact that the entire range in $\delta_{12}$ is less than $0.2$~mag for GY~314, we are unable to confidently claim that this extinction law accurately reproduces the data for this YSO. We thus see no evidence for (or against) extinction-induced mid-IR variability in any of our targets from their light curves alone; in \S \ref{sec:extinct}, we utilize our spectral data to investigate this possibility in more detail.

\begin{deluxetable*}{c | c c c c c c c c}
\tablecolumns{9}
\tablewidth{0pt}
\tablecaption{Mid-IR and NIR variability indices}
\tablehead{
\colhead{Object} &	\colhead{$\Delta[3.6]$} &	\colhead{$\Delta[4.5]$} &	\colhead{$\langle \mathrm{EW}_{Br \gamma} \rangle $\tablenotemark{a}} &	\colhead{min. EW$_{Br \gamma}$\tablenotemark{a}} &	\colhead{max. EW$_{Br \gamma}$\tablenotemark{a}} &	\colhead{$\langle \mathrm{EW}_{Pa \beta} \rangle $\tablenotemark{a}} &	\colhead{$\langle r_H \rangle$} &	\colhead{$\langle r_K \rangle$} \\
\colhead{} &	\colhead{(mag)} &	\colhead{(mag)} &	\colhead{(-\AA)} &	\colhead{(-\AA)} &	\colhead{(-\AA)} &	\colhead{(-\AA)} &	\colhead{} &	\colhead{}
}
\startdata
GY 292 & 0.07 & 0.07 & $4.9 \pm 1.0$ & $1.7 \pm 0.6$ & $11.6 \pm 5.1$ & $5.5 \pm 1.1$ & 1.4 & 2.5 \\
GY 308 & 0.18 & \nodata & $3.7 \pm 0.6$ & $1.4 \pm 0.5$ & $6.4 \pm 1.6$ & $4.4 \pm 0.7$ & 0.9 & 1.2 \\
GY 314 & 0.21 & 0.17 & $1.6 \pm 0.2$ & $0.8 \pm 0.8$ & $2.5 \pm 0.8$ & $0.7 \pm 0.3$ & 0.5 & 0.9 \\
YLW 15a & 0.12 & 0.09 & \nodata & \nodata & \nodata & \nodata & \nodata & \nodata \\
YLW 16b & 0.44 & 0.42 & \nodata &\nodata & \nodata & \nodata & \nodata & \nodata \\ 
\enddata
\tablenotetext{a}{Veiling-corrected EWs}
\label{tab:mastertable}
\end{deluxetable*}

\subsection{Spectral Measurements}

A few key parameters dominate the morphology of a YSO's NIR spectrum: (1) the effective temperature of the stellar photosphere ({\Teff}), primarily diagnosed by the presence and relative strength of prominent (Na, CO, Ca, H$_2$O, etc.) absorption features; (2) the amount of veiling flux, diagnosed by the absolute depth of photospheric features relative to a template of similar spectral type; (3) extinction, diagnosed by the broadband slope of the spectral continuum; and, (4) the mass accretion rate, diagnosed by the strength of prominent emission lines such as Paschen $\beta$ ({\pab}, \ion{H}{1}~$\lambda12818$) and Brackett $\gamma$ ({\brg}, \ion{H}{1}~$\lambda 21661$). In this section, we follow previously employed techniques to infer these physical parameters for our YSO targets from moderate resolution NIR spectra \citep[e.g., LR99;][]{Winston09,Covey10} to assess which, if any, physical properties correlate strongly with mid-IR variability.

\subsubsection{Spectral Types}
\label{sec:spt}

Visual inspection of strong, temperature sensitive absorption lines in Figure~\ref{fig:spectra} does not reveal any evidence for significant changes in the photospheric temperature of these YSOs. Specifically, the relative strengths of nearby temperature sensitive lines (e.g., the 2.21~$\mu$m \ion{Na}{1} \& 2.26~$\mu$m \ion{Ca}{1} doublets) appear consistent from epoch to epoch, modulo small changes in overall depth due to modest changes in the star's continuum veiling. Given the evident stability in each YSO's photospheric temperature, we have estimated each star's photospheric spectral type from the spectrum obtained on April 1st; we consider potential veiling and extinction variations in more detail in \S\S~\ref{sec:VeilVar} and \ref{sec:extinct}, respectively. We estimated each star's spectral type using the algorithm developed and presented in full by \cite{Covey10}; we refer the interested reader to that work for a complete description of the spectral analysis procedure, and simply summarize here the most pertinent details.

Spectral type and veiling estimates are derived by comparing the absolute and relative strengths of temperature sensitive spectral features (e.g., Na, Ca, CO, Mg, H$_2$O; for a full list see Tables 3 and 4 of \citealp{Covey10}) measured from spectral standards observed by \cite{LW00}, \cite{CRV05} and \cite{Covey10}. These standards span a wide range of effective temperatures, and include giants and dwarfs, as well as pre-main sequence stars in the TW Hya association, enabling an estimation of the target's surface gravity as well as its {\Teff}. Estimates of the extinction toward each YSO photosphere are determined by comparing its NIR spectral slope, as quantified by its spectrally integrated $J-H$ and $H-K_s$ colors, to the colors expected for an artificially reddened and veiled spectral standard. We computed these spectrally integrated colors by convolving each SpeX spectrum with standard 2MASS filter curves and integrating the total transmitted flux, using routines originally developed by \cite{Covey07}. The spectrally integrated $J-H$ and $H-K_s$ colors derived from each target spectrum are reported in Table \ref{tab:obslog}.

We determined the spectral type, veiling and extinction that provide the best overall fit to each YSO's NIR spectrum using an iterative technique to refine the initial values we assume for each parameter. We began by adopting a fiducial K7 spectral type for each YSO, a common spectral type for typical T~Tauri stars. Comparing the overall spectral slope (as parameterized by the spectrally integrated $J-K_s$ color) and strengths of prominent absorption features in the target spectrum to those of a pre-main sequence K7 spectral standard, we derived initial estimates of the YSO's extinction and veiling, respectively. We then synthetically reddened and veiled the grid of spectral standards to match these initial estimates, and compared the strengths of the temperature and gravity-sensitive absorption features in the YSO's spectrum to those measured from the synthetically reddened and veiled
standards. The un-reddened, un-veiled spectrum of the new {\Teff} standard could then be used to produce updated estimates of the source's veiling and extinction, which in turn could be re-applied to the grid of standard spectra.  Repeating this process several times was sufficient to identify a self-consistent set of extinction (which we report in the $H$~band as {\AH}), veiling, and spectral type estimates that reproduced each targetÕs spectral morphology on both large and small spectral scales. We compute {\Teff} from spectral type using the scale derived by LR99, and report the spectral type (SpT), {\Teff}, and {\AH} for each of our targets in Table~\ref{tab:mdot} alongside our derived mass accretion rates (see \S \ref{sec:EmLineVar}).

Previous tests of this algorithm have identified a characteristic uncertainty in spectral type estimates of $\sigma$=$\pm$1 subclass; to incorporate the possibility of veiling and/or extinction variability, we conservatively adopt an uncertainty of $\sigma$=$\pm$2 subclasses in our spectral type determinations. Estimating spectral types for YSOs, particularly at the youngest evolutionary stages, is challenging, and there are discrepancies between the previously published results for these YSOs. LR99 used moderate-resolution $K$~band spectra to derive spectral types of K8, K8, and M0 for GY 292, 308, and 314, respectively, while \cite{Gatti06} determined somewhat earlier values of K3, K5, and K5 from low-resolution $J$~band spectra; both authors report uncertainties of $\sim$3 subclasses. Our results of K5, M0, and M2, respectively, agree with the types LR99 derived for all three YSOs to within our $\pm$2 subclass uncertainty, but are in somewhat worse agreement with those derived by \cite{Gatti06}.

We suspect that the agreement between the types derived here and by LR99, and the systematic offset towards earlier types in the analysis of \cite{Gatti06} is primarily due to a key difference in the treatment of veiling in these works. The types inferred here and by LR99 explicitly allow for line dilution by veiling, whereas \citet{Gatti06} assume that the $J$~band is veiling-free, and derive types based on absolute line strengths. If these YSOs possess non-negligble $J$~band veiling, as many CTTSs do \citep{Cieza05}, their $J$~band photospheric lines will appear systematically weaker than they would in the absence of veiling. Many atomic features in NIR stellar spectra, including those analyzed by \cite{Gatti06}, weaken towards hotter {\Teff}s, such that the dilution of a YSO's line strengths by veiling flux can be misdiagnosed as indicative of a warmer photosphere, and thus an earlier spectral type. The analysis employed here, as well as by LR99, assigns spectral types based on ratios of lines with somewhat divergent temperature dependences, and thus minimizes (but does not completely remove) this degeneracy between {\Teff} and veiling flux; once the YSO's spectral type has been established by these line ratios, the absolute line depths can be used to infer the presence of additional veiling flux. Furthermore, the trend between absolute line strength and {\Teff} is rather shallow for the $J$~band features used by \cite{Gatti06} to infer spectral types, such that the intrinsic scatter between line strengths at a particular {\Teff} is comparable to the magnitude of the slow trend over a full spectral class. If the \cite{Gatti06} uncertainties were increased to reflect this additional degeneracy, it would better reconcile their results with ours.

The degeneracy between {\Teff} and veiling discussed above is sufficiently challenging that veiling remains the most poorly constrained parameter by our iterative spectral fitting technique. Reasonable by-eye fits to the object spectra can be obtained with veilings that differ by as much as 50\%, after allowing for a slight modification in {\Teff} (which we incorporate into our spectral type uncertainty) to offset the change in absolute line depth. We therefore describe in the next section a separate, objective measurement of potential veiling variations in each YSO's spectrum once an appropriate spectral template has been identified.

We estimate the uncertainty in the extinction {\AH} derived from our spectral analysis by deriving alternative {\AH} estimates from dereddening the published 2MASS magnitudes for our YSOs to the CTTS locus~\citep{Meyer97}. Comparing these 2MASS-based {\AH} values to those obtained using our spectrally integrated $JHK_s$ magnitudes reveals that the two estimates are consistent to $\sim$0.3 magnitudes in {\AH}, and the differences may be partially due to long-term photometric variability. However, as further discussed in \S \ref{sec:diskemission}, the NIR magnitudes of these YSOs are relatively stable between the measurements obtained by \cite{Barsony97} and the 2MASS survey, so we conservatively adopt the full $0.3$~mag as the uncertainty in {\AH}. For consistency with the line luminosity calculations in section \ref{sec:EmLineVar}, the {\AH} values listed in Table \ref{tab:mdot} are those determined from their published 2MASS magnitudes.

\subsubsection{Veiling Measurements}
\label{sec:VeilVar}

As mentioned in the previous sections, young stars often exhibit `veiling', an excess of near-infrared emission which presumably arises from the inner edge of a cool circumstellar disk. As a result, photospheric absorption lines in YSO spectra can be significantly weaker than in field stars of similar {\Teff} when the spectra are normalized to the continuum. In this section, we analyze our multi-epoch NIR spectra to directly measure the time evolution of NIR veiling in our spectroscopic targets.

To begin, we parameterize veiling in terms of an absorption feature's equivalent width (EW), a continuum-independent observable that is robust against absolute flux calibration uncertainties. If we define $I_{\mathrm{line}}$ to be the total absorbed flux in a photospheric line, the EW of a spectral feature in an \textit{unveiled} star's spectrum is given by EW$_{\mathrm{unveiled}} = I_{\mathrm{line}}/\langle F_* \rangle$, where $\langle F_* \rangle$ is the continuum flux at the wavelength of the feature. In contrast, the continuum flux in a star with the same {\Teff} but nonzero excess emission from the disk is $F_{\mathrm{cont}} = F_{\mathrm{excess}} + F_*$, where $F_{\mathrm{excess}}$ is the disk emission at that wavelength. Thus the EW of the exact same spectral line in the veiled star will instead be EW$_{\mathrm{veiled}} = I_{\mathrm{line}}/\langle F_* + F_{\mathrm{excess}} \rangle$. A target star's veiling $r_{\lambda}$ can therefore be calculated from the equivalent widths of an absorption line in the spectrum of the target and in an unveiled spectral standard of the same spectral type. Using the above relations with the fact that veiling is defined as $r_{\lambda} \equiv F_{\mathrm{excess}}/F_*$ yields the relation:
\begin{eqnarray}
r_\lambda = \frac{EW_{\mathrm{unveiled}}}{EW_{\mathrm{veiled}}} - 1.
\label{eqn:eqwalt}
\end{eqnarray}
To estimate each target's veiling at all epochs and a range of wavelengths, we measured the EWs of several prominent $H$ and $K$ band photospheric absorption features in each spectrum, selecting consistent line centers and widths for each feature so as to avoid contamination from neighboring features while still capturing the full line profile at each epoch. We also chose the sizes and widths of continuum regions on either side of each line to avoid nearby lines. We present the complete list of features measured in each target (including the emission features discussed in \S \ref{sec:EmLineVar}) in Table~\ref{tab:lineparams} along with our chosen line and continuum parameters. We also measured the EWs of each absorption feature in standard stars of the appropriate spectral type, then used Equation~\ref{eqn:eqwalt} to infer veiling for each of our targets at each epoch. Table~\ref{tab:veiling} presents the minimum, maximum and mean veilings, as well as the dispersion in the veiling time series, for each measured feature\footnote[3]{Our measured $K$~band veilings $r_K$ are systematically higher than those of LR99 by a factor of $\sim2$. To investigate this discrepancy, we compared veiling measurements obtained using different combinations of target spectra and measurement methodology. Specifically, we compared veiling measurements obtained by: (1) applying the methodology described above to the LR99 spectra, and (2) applying the LR99 methodology to our SpeX spectra (K. Luhman, private communication). The veilings we derive from these tests are consistent with those in Table~\ref{tab:veiling} to well within our uncertainties. This suggests the most likely cause for the disagreement between our measurements and those of LR99 lies in the choice of spectral standards used to define an unveiled photosphere: the veilings listed in Table~\ref{tab:veiling} and via these tests were all derived using SpeX spectra of dwarf stars as proxies for unveiled photospheres, whereas LR99 measured veiling with respect to spectral standards generated by averaging spectral of dwarf and giant stars.}.

\begin{deluxetable*}{c | ll | llll}
\tablecolumns{7}
\tablewidth{0pt}
\tablecaption{Spectral Line Parameters}
\tablehead{
\colhead{Line} &	\colhead{Line Center} &	\colhead{Line Width} &	\colhead{Cont. 1\tablenotemark{a}} &
\colhead{Cont. 1\tablenotemark{a}} &	\colhead{Cont. 2\tablenotemark{a}} &
\colhead{Cont. 2\tablenotemark{a}} \\
\colhead{Name} &	\colhead{(\micron)} &	\colhead{(\micron)} &	\colhead{Center (\micron)} &
\colhead{Width (\micron)} &	\colhead{Center (\micron)} &	\colhead{Width (\micron)}
}
\startdata
Mg I	& 1.5037\tablenotemark{b}	& 0.005 & 1.4982 & 0.0028 & 1.5091 & 0.0028 \\
Al I	& 1.6740 					& 0.006 & 1.6696 & 0.0020 & 1.6781 & 0.0020 \\
Mg I	& 1.7110					& 0.003 & 1.7044 & 0.0032 & 1.7177 & 0.0032 \\
Ca I	& 1.9820\tablenotemark{c}	& 0.013 & 1.9681 & 0.0027 & 1.9972 & 0.0027 \\
Na I	& 2.2071\tablenotemark{b}	& 0.006 & 2.1991 & 0.0064 & 2.2152 & 0.0064 \\
Ca I	& 2.2640\tablenotemark{b}	& 0.008 & 2.2565 & 0.0043 & 2.2715 & 0.0043 \\
Mg I	& 2.2812					& 0.003 & 2.2755 & 0.0027 & 2.2868 & 0.0027 \\
\tableline
He$_{\mathrm{blue}}$\tablenotemark{d} 	& 1.08220 & 0.0014 & 1.0803 & 0.0013 & 1.0859 & 0.0013 \\
He$_{\mathrm{red}}$\tablenotemark{e}	& 1.08335 & 0.0009 & 1.0800 & 0.0013 & 1.0859 & 0.0013 \\
\pag & 1.0940 & 0.0025 	& 1.0908 & 0.0019 & 1.0972 & 0.0019 \\
\pab & 1.2817 & 0.0025	& 1.2782 & 0.0023 & 1.2853 & 0.0023 \\
\brg & 2.1655 & 0.006 	& 2.1586 & 0.0038 & 2.1725 & 0.0038
\enddata
\tablenotetext{1}{Continuum regions were chosen so as to include the entirety of the spectral line at all epochs for all YSOs while minimizing contamination from neighboring features.}
\tablenotetext{2}{Doublet; the reported line center is the mean of the two individual line centers}
\tablenotetext{3}{Triplet; the reported line center is the mean of the three individual line centers}
\tablenotetext{4}{The portion of the He line just blueward of the star's rest frame velocity (see \S \ref{sec:HeI})}
\tablenotetext{5}{The portion of the He line just redward of the star's rest frame velocity (see \S \ref{sec:HeI})}
\label{tab:lineparams}
\end{deluxetable*}

\begin{deluxetable*}{l | r | c c c c | c c c c | c c c c}
\tablecolumns{14}
\tablewidth{0pc}
\tablecaption{Veiling measured from photospheric absorption lines}
\tablehead{
\colhead{} & \colhead{} & \multicolumn{4}{c}{GY 292} & \multicolumn{4}{c}{GY 308} & \multicolumn{4}{c}{GY 314} \\
\tableline
\colhead{line} & \colhead{$\lambda (\micron)$} &
\colhead{$r_{\lambda,\mathrm{min}}$} & \colhead{$r_{\lambda,\mathrm{max}}$} & \colhead{$\langle r_\lambda \rangle$} & \colhead{$ \langle \sigma_r \rangle $} &
\colhead{$r_{\lambda,\mathrm{min}}$} & \colhead{$r_{\lambda,\mathrm{max}}$} & \colhead{$\langle r_\lambda \rangle$} & \colhead{$ \langle \sigma_r \rangle $} &
\colhead{$r_{\lambda,\mathrm{min}}$} & \colhead{$r_{\lambda,\mathrm{max}}$} & \colhead{$\langle r_\lambda \rangle$} & \colhead{$ \langle \sigma_r \rangle $}}
\startdata
Mg I &	1.5037 &	0.58	&	1.00 &	0.81	&	0.15 &
	0.26 & 0.54 & 0.36 & 0.09 &
	0.00 & 0.16 & 0.02 & 0.08 \\
Al I & 	1.6700 &	0.96 &	1.84 &	1.46 &	0.32 &
	0.40 & 1.49 & 0.95 & 0.35 &
	0.42 & 1.10 & 0.68 & 0.26 \\
Mg I & 1.7110 &	0.87 &	2.39 &	1.40 &	0.43 &
	0.66 & 1.11 & 0.86 & 0.18 &
	0.33 & 0.73 & 0.52 & 0.14 \\
Ca I & 1.9820 &	1.13 &	3.23 &	1.99 &	0.71 &
	0.78 & 1.58 & 1.05 & 0.23 &
	0.56 & 1.40 & 0.97 & 0.31 \\
Na I & 2.2071 &	1.64 &	3.44 &	2.49 &	0.68 &
	0.80 & 1.64 & 1.22 & 0.27 &
	0.69 & 1.36 & 1.01 & 0.24 \\
Ca I & 2.2640 &	1.58 &	3.81 &	2.54 &	0.80 &
	0.80 & 1.50 & 1.14 & 0.24 &
	0.59 & 1.18 & 0.82 & 0.22 \\
Mg I & 2.2820 &	0.90 &	3.80 &	1.89 &	0.95 &
	0.56 & 1.53 & 0.91 & 0.32 &
	0.27 & 0.62 & 0.41 & 0.11 \\
\enddata
\label{tab:veiling}
\end{deluxetable*}

\begin{figure}
\includegraphics[width=0.8\linewidth,angle=90]{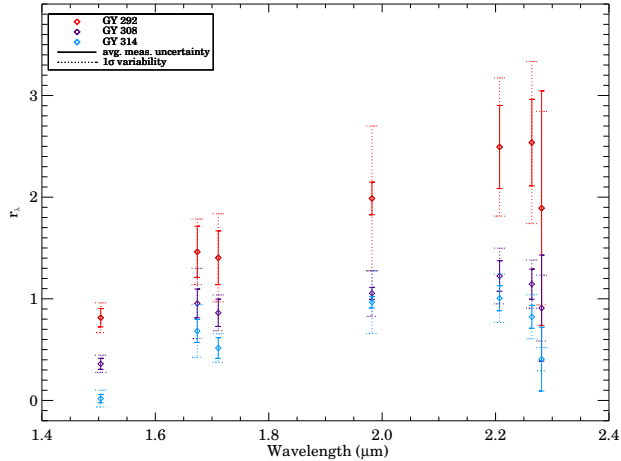}
\caption{\small{Veiling as a function of wavelength, as inferred from individual spectral features in GY 292, 308, and 314. The solid error bars show the average measurement uncertainties, while the dotted error bars represent the $1\sigma$ variability in veiling across all epochs. Veiling increases as a function of wavelength across the $J$ and $H$ bands in each YSO's spectrum, then flattens across the $K$ band. The ordering of objects by veiling strength is identical across the NIR, with GY 292 and 314 exhibiting the largest and smallest values, respectively. The \ion{Mg}{1}~$\lambda 22814$ feature is very weak in our targets, and thus the uncertainties in veiling inferred from this line are large.}}
\label{fig:allveil}
\end{figure}

Emission from a circumstellar disk typically increases or remains relatively constant through the NIR and mid-IR\footnote[4]{Disk SEDs are in truth exceptionally complicated in the infrared; the exact details depend on a host of factors including dust grain size distribution, grain growth properties, disk geometry, viewing angle, stellar irradiation, and interstellar UV field. Our goal is only to empirically measure the wavelength dependence of veiling in our targets without attempting to model the disk in any great detail.} \citep[e.g.,][]{Dalessio06}, in contrast to photospheric emission which is decreasing along the power-law Rayleigh-Jeans blackbody tail at these wavelengths. Thus one would expect veiling in YSOs to increase across the NIR. To confirm this scenario, we computed time-averaged line-by-line veiling measurements for each YSO, which are presented as a function of wavelength in Figure~\ref{fig:allveil}; solid error bars show average measurement errors, while dotted error bars represent the formal standard deviation of the veiling time series. Veiling does indeed increases with wavelength across the $JHK$ bands in all three YSOs, although there is some evidence for a decrease in slope around $2.2~\mu$m. We note, however, that the \ion{Mg}{1}~$\lambda 22814$ feature was particularly weak in our targets, leading to high formal uncertainties on veiling measurements at $2.28~\mu$m, and thus we do not accord any physical significance to the apparent systematic decrease in veiling at the upper end of the $K$ band.

To characterize the typical veiling levels in the $H$ and $K$ bands and their potential time variability for comparison with our IRAC light curves, we select the strongest features near the center of each band (and therefore best separated from telluric H$_2$O features), and define the veiling inferred from these lines as $r_H$ and $r_K$. In particular, we define $r_H=r($\ion{Mg}{1}~$\lambda17110)$ and $r_K=[r($\ion{Na}{1}~$\lambda22071) + r($\ion{Ca}{1}~$\lambda22640)]/2$. We present time-averaged veilings $r_H$ and $r_K$ in Table~\ref{tab:mastertable}, and visually show the time evolution of $r_K$ alongside the other NIR and mid-IR indices in Figure~\ref{fig:master}. Veiling in all three YSOs appears to vary only modestly over our observation window (note that variations in $r_K$ have been suppressed by a factor of two in the figure for readability).

We estimated uncertainties in EW measurements through a Monte Carlo approach devised for SpeX NIR spectra (see \citealp{Muirhead11} for a complete description). In summary, artificial noise was added to each YSO spectrum, with the noise drawn from a Gaussian probability distribution function (PDF) having a width equal to the flux uncertainty estimated for each pixel by the SpeXtool pipeline. The EW of each spectral feature was then measured from this artificially noised spectrum. This process was independently repeated 1000 times for each EW measurement; the error in a given EW measurement is then taken to be the standard deviation of all the simulated EW measurements for that feature. To independently estimate the effect of a potentially misplaced continuum, we separately computed EWs of several emission and absorption features in our fiducial April 1 spectrum while artificially varying the continuum level by $\pm 2\%$ (the approximate range over which the by-eye placement appeared reasonable). Over this range of continuum placement levels, EWs varied by an average of $\sim 0.10$~\AA, so we added this value in quadrature with the Monte Carlo-determined error to determine the total EW uncertainty, which is then formally propagated into our veiling and mass accretion rate errors.

We note that in all three YSOs, the photospheric features seen in each spectrum appear to be systematically weaker in the last six epochs as compared with the first four, leading to relatively higher veiling measurements at the end of the monitoring program. This could potentially be due to the fact that different standard stars were used for telluric corrections in the first four epochs as compared to the last six. To explore this possibility, we consider the relative strengths of absorption features in target spectra reduced with both telluric standards, with all spectra obtained at comparable airmasses on April~16 (MJD~2455303). Specifically, we measure the EW of the full suite of absorption features listed in Table~\ref{tab:lineparams} in each of two alternate reductions of both GY~292 and GY~314, one reduced with HD~145188 (the standard used for the majority of the reductions for the last six epochs), and the other reduced with HD~145127 (one of the standards used for the first four epochs). We find that EWs measured from spectra reduced with HD~145188 are systematically lower (in all lines and both YSOs) by $\sim$0.1--0.5~{\AA} (1--2$\sigma$) than those reduced with HD~145127. As Equation~\ref{eqn:eqwalt} shows, a 0.1--0.2 {\AA} change in the EW of a feature in the standard star's spectrum (EW$_{\mathrm{unveiled}}$) leads to a 15--30\% difference in measured veiling in these YSOs. There are no obviously visible photospheric absorption features in the standard star spectra, but we are unable to rule them out below the $\sim$0.5~{\AA} level due to the large number of nearby telluric features. There is also a substantive difference in the characteristic airmasses of the observations obtained during the first four epochs ($\sec z = 1.4$ to 1.5) compared to the last six epochs ($\sec z = 1.6$ to 2.2), so imperfect subtraction of atmospheric lines at high airmasses could have also contributed to this effect. Nonetheless, the IRAC observation window only overlaps the last six spectroscopic epochs, and thus our primary conclusions (which are based on relative, direct temporal comparison of mid-IR light curves and veiling) are entirely unaffected by this potential systematic effect. In addition, hydrogen features (e.g. {\brg} and {\pab}) are well-subtracted in the telluric calibration, so EW measurements of these lines are insensitive to the choice of standard star.

\subsubsection{Accretion \& Outflow Line Diagnostics}
\label{sec:EmLineVar}

\cite{MHC98} demonstrated that the luminosities of NIR hydrogen emission lines (specifically, {\brg} and {\pab}) emitted by young, low-mass (0.2--0.8 \Msun) stars correlate well with their mass accretion rates as derived from ultraviolet (UV) observations. These authors also show a slightly weaker correlation between the equivalent widths (EW) of these lines and YSO mass accretion rates; the advantage of using EWs for our analysis is that they reflect the intrinsic strength of the line relative to the local continuum, and thus avoid errors associated with a spectrum's absolute flux calibration. As our spectra are subject to substantial uncertainties in their absolute flux calibration, {\brg} and {\pab} EWs provide the most reliable means of investigating changes in our YSO targets' accretion rates over time; due to the superior S/N achieved in our $K$ band spectra, we selected {\brg} in particular for detailed time series analysis of the mass accretion rate.

Our measurements of EW({\brg}) at each epoch are presented visually alongside our other NIR and mid-IR quantities in Figure~\ref{fig:master}; variability in this feature for all three YSOs is clearly evident (note that variations have been suppressed by a factor of two in the figure for readability). Combining our measurements to study the characteristic accretion activity of our YSO targets over the entire 40-day observation window with improved statistical power, we also computed time-averaged equivalent widths $\langle$EW({\brg})$\rangle$ for each YSO, which we present in Table~\ref{tab:mastertable}. Both individual and time-averaged measurements have been veiling-corrected to remove the disk contribution to the continuum that would otherwise bias EWs downward in more heavily veiled spectra. Note that the uncertainty reported on $\langle$EW({\brg})$\rangle$ is the formal uncertainty of the mean, defined as $\sigma_{\mathrm{m}} \equiv \sigma/\sqrt{N}$, where $\sigma$ is the standard deviation of the data and $N$ is the number of measurements.

To place each YSO's characteristic mass accretion rate on an absolute physical scale, we converted time-averaged {\brg} EWs into line luminosities by multiplying by the continuum flux level inferred from the 2MASS $K_s$~magnitude, after correcting for distance (d$\sim 120\pm5$~pc; \citealp{Loinard08}) and extinction (using our derived $A_H$ and the extinction law of~\citealp{Fitzpatrick99}). The total accretion luminosity {\Lacc} of each object can then be derived from the empirical correlations between {\Lacc} and {\brg} line luminosity first presented by \cite{MHC98} and later refined by \citet[][see their equations 3 \& 4]{NTR06}. The mass accretion rate is then given by $\Macc = \Lacc R_* / (GM_*)$. For this calculation, we estimated each YSO's radius by interpolating the {\Teff} estimates inferred from our spectral typing procedure onto the pre-main sequence evolutionary tracks computed by \cite{Baraffe98}, after adopting a characteristic age of 3.5~Myr \citep[the median age of $\rho$~Oph YSOs inferred from this model grid;][]{Covey10}; for each YSO, we adopted the mass reported by \cite{NTR06}. GY~292 and GY~308 also exhibit strong {\pab} emission lines (Figure~\ref{fig:spectra}), and we follow the above procedure using time-averaged {\pab} EW and 2MASS $J$~magnitude to derive {\Macc} from the {\pab} line in these two objects as well. We summarize the \ion{H}{1} line properties and derived accretion rates in Table~\ref{tab:mdot}; the implied time-averaged mass accretion rates range from about 10$^{-8.5}$ M$_{\odot}$/yr for GY 292, to 10$^{-9.2}$ M$_{\odot}$/yr for GY 314. The combination of uncertainties in $A_H$ ($\sim$ 0.3~mag), distance ($\sim 4\%$), and the EW measurement ($\sim$10--20\%) lead to a formal uncertainty of $\sim30\%$ in {\brg} and {\pab} line luminosity. Errors in accretion luminosity are thus dominated by the scatter in the $L_{\mathrm{acc}}-$line luminosity relation, which \cite{Natta04} place as a factor of 2 to 3. For {\Macc}, uncertainties in mass ($\sim60\%$) and radius ($\sim40\%$) also contribute, and we conservatively estimate a total error of a factor of 3 (equivalent to $\sim 0.5$~dex) in {\Macc}. Despite the relatively large uncertainties, we see a measurable difference in time-averaged mass accretion rages amongst our three YSO targets; we will discuss these results in the context of YSO evolution in \S \ref{sec:ages}.

\begin{deluxetable*}{l | ccc | cc | cc | cc}
\tablecolumns{10}
\tablewidth{0pt}
\tablecaption{Mass Accretion Rates}
\tablehead{
\colhead{Object}	& \colhead{SpT}	& \colhead{\Teff}\tablenotemark{a}	& \colhead{$A_H$}	&
\colhead{$L$({\brg})}	& \colhead{$L$({\pab})}	& \colhead{$L_{\mathrm{acc}}$({\brg})}	 & \colhead{$L_{\mathrm{acc}}$({\pab})}	& \colhead{$M_{\mathrm{acc}}$({\brg})}	& \colhead{$M_{\mathrm{acc}}$({\pab})} \\
\colhead{}			& \colhead{}	& \colhead{\small{(K)}}	& \colhead{\small{(mag)}}	&
\multicolumn{2}{c}{\small{($\log L/\Lsun$)}} & \multicolumn{2}{c}{\small{($\log L/\Lsun$)}} &
\multicolumn{2}{c}{\small{($\log M/\Msun$/y)}}
}
\startdata
GY 292	& K5	& 4350	& 2.21	& -4.11	& -3.53	& -0.80	& -0.80	& -8.47	& -8.47 \\
GY 308 	& M0	& 3850	& 1.86	& -4.30	& -3.74	& -0.97	& -1.08	& -8.55 	& -8.67 \\
GY 314	& M2	& 3514	& 1.12	& -4.89	& \nodata	& -1.50	& \nodata	& -9.19	& \nodata
\enddata
\tablenotetext{a}{Computed from spectral type using the scale derived by LR99}
\label{tab:mdot}
\end{deluxetable*}

\begin{figure*}
\vspace{-10 mm}
\begin{tabular}{cc}
\includegraphics[width=0.5\linewidth]{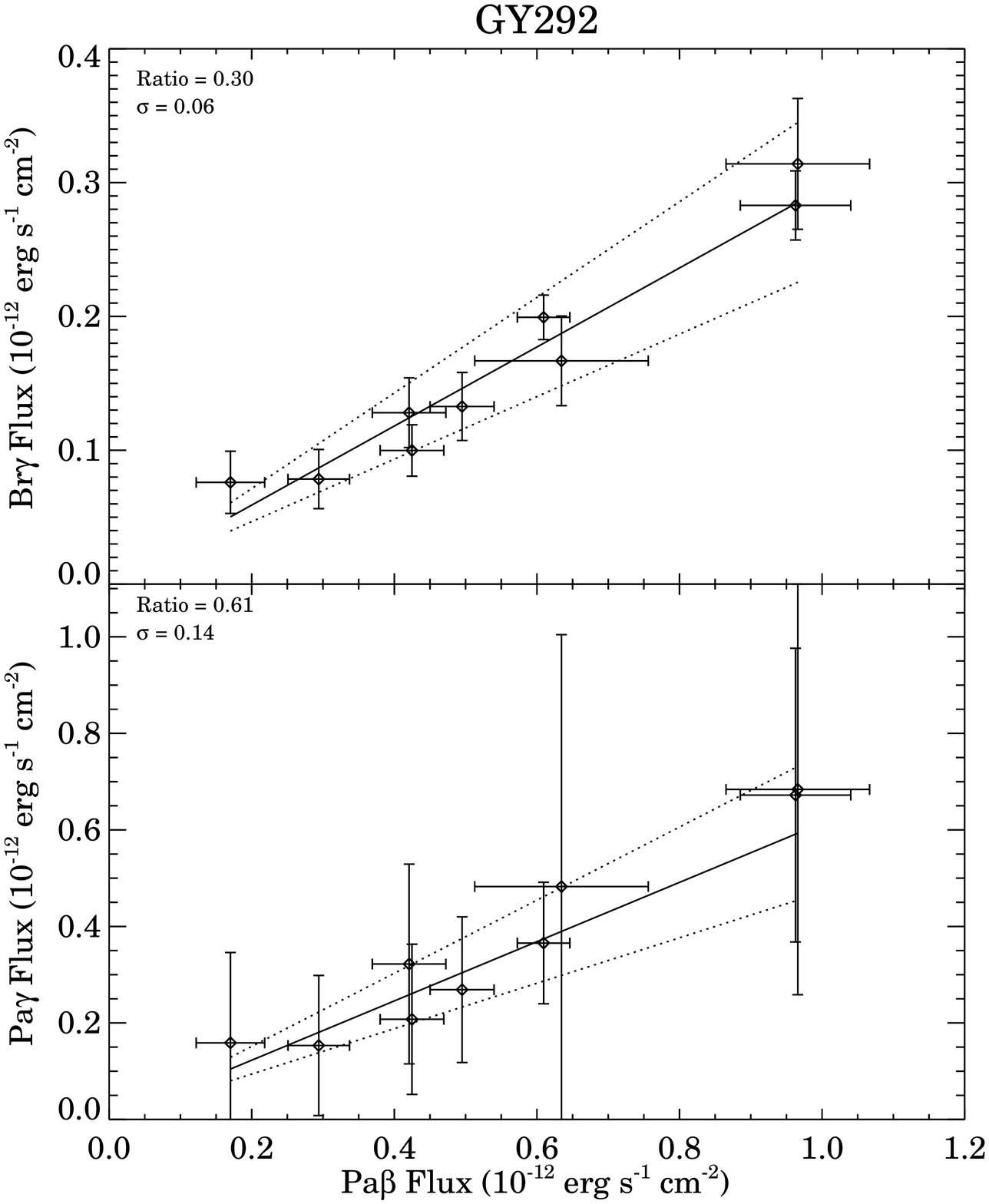} &
\includegraphics[width=0.5\linewidth]{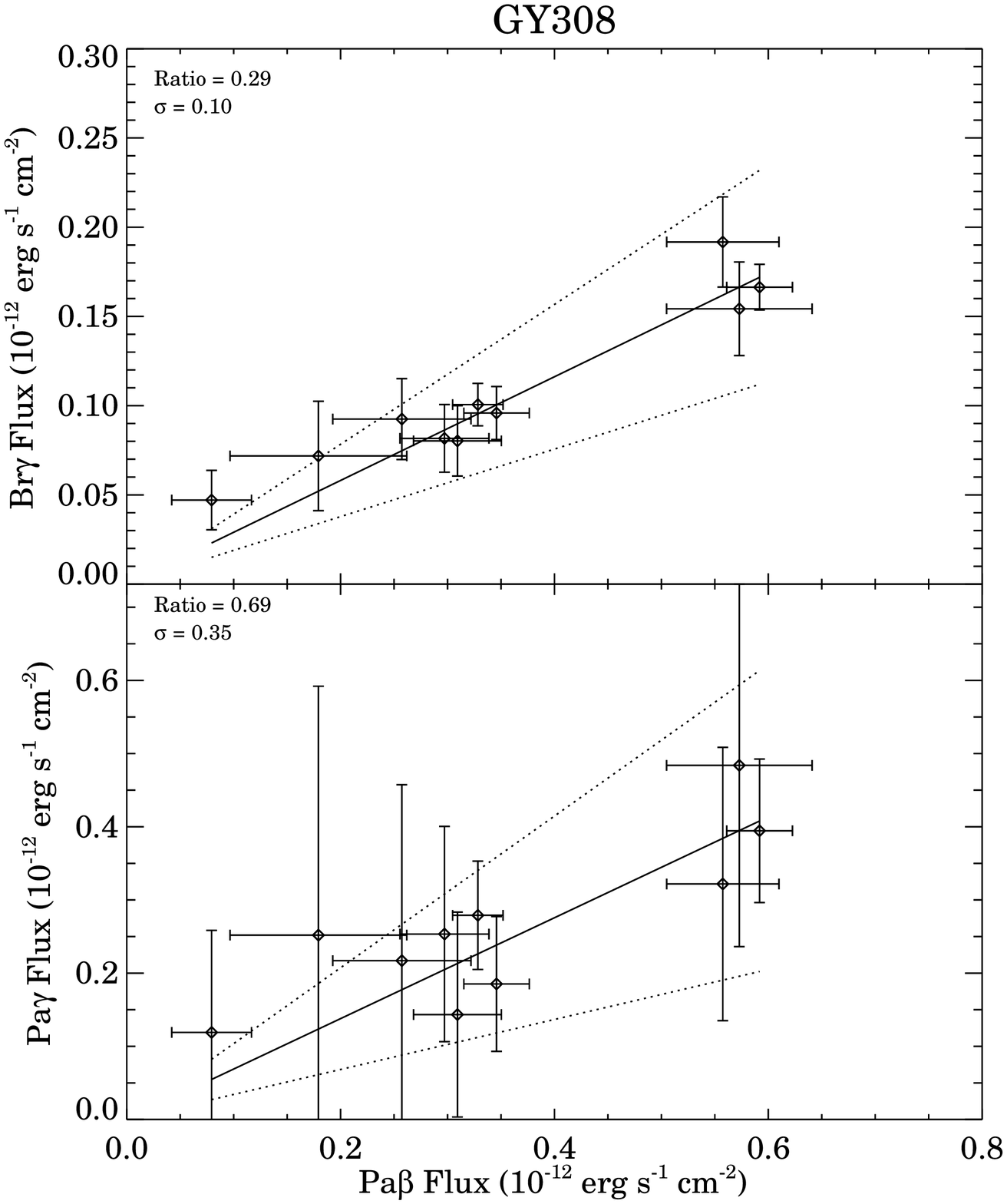} \\
\end{tabular}
\caption{\small{{\brg}/{\pab} (top panels) and Pa$\gamma$/{\pab} (bottom panels) line ratios for GY~292 and GY~308, the two YSOs with reliably detected {\pab} and Pa$\gamma$ emission. The solid line in each panel has slope equal to the derived line ratio, which we compute as the weighted average ratio of the data; $1\sigma$ intervals are plotted as dotted lines. Our {\brg}/{\pab} line ratios are consistent with {\brg}/{\pab}$ \sim 0.25$, the value derived by \cite{Bary08} for a large sample of YSOs. This line ratio can be reproduced by models assuming Case B recombination theory for a wide range of temperatures and densities. There is no compelling evidence for variability in the either line ratio.}}
\label{fig:lineratios}
\end{figure*}

For the two YSOs with detected {\pab} and Pa$\gamma$ emission (GY~292 and GY~308), we can compute {\brg}/{\pab} and {\pag}/{\pab} line ratios to look for consistency with Case~B recombination theory, which assumes that the emitting gas is optically thick to Lyman series photons and optically thin to all other \ion{H}{1} transitions~\citep[e.g.,][]{SH95}. Using the extinctions ($A_H$) derived in \S \ref{sec:spt}, we dereddened each YSO's 2MASS $J$ and $K$ band magnitudes using an $R_V = 3.1$ \cite{Fitzpatrick99} extinction law. Converting these dereddened 2MASS magnitudes into estimates of each YSO's de-reddened continuum flux densities, we converted our {\brg}, {\pab}, and Pa$\gamma$ EWs into line fluxes. In Figure~\ref{fig:lineratios} we plot line fluxes $F_{\mathrm{Br}\gamma}$ and $F_{\mathrm{Pa} \gamma}$ as functions of $F_{\mathrm{Pa}\beta}$ for both GY~292 and GY~308. Computing the weighted average ratios from the data, we find {\brg}/{\pab} line ratios of $0.30 \pm 0.07$ and $0.29 \pm 0.10$, and Pa$\gamma/$Pa$\beta$ line ratios of $0.61 \pm 0.14$ and $0.69 \pm 0.35$ for GY~292 and GY~308, respectively. Line ratios are consistent between the two YSOs, and there is no evidence for significant variability in either line ratio to within our measurement uncertainties. Note that the 1$\sigma$ errors listed above represent the standard deviations of the line ratio time series, and include neither measurement uncertainties nor the additional error introduced by assuming that the 2MASS magnitudes represent accurate continuum fluxes. The uncertainties in the calculated Pa$\gamma$/{\pab} line ratios are likely to be even higher due to the fact that we use a single value for the $J$~band continuum (the 2MASS measurement) to compute both line fluxes despite their relatively large wavelength separation. Our derived {\brg}/{\pab} results are consistent with the line ratio of $\sim 0.25$ measured in a sample of YSOs in Taurus by \cite{Bary08}, as well as with the predictions of Case B recombination for a wide range of realistic electron temperatures and densities~\citep{SH95}. However, the above complications, combined with the fact that our S/N is inadequate to reliably measure additional higher-$n$ transitions in either the Paschen or Brackett series, render us unable to directly constrain the electron temperature or density in the emitting regions of these YSOs using our observations.

The \ion{He}{1}~$\lambda 10830$ line has also proven to be an informative probe of outflow and accretion activity in young stars (e.g., \citealp{Edwards06,Fischer08}). To ensure the reliability of the kinematic content of the rich \ion{He}{1} line profile, we have compensated for any potential remaining wavelength calibration errors in our spectra by adjusting the {\pab} line center to the {\pab} rest wavelength at all epochs using the following procedure. First, we measured the center of the {\pab} feature in the spectra of GY~292 and GY~308 (we excluded GY~314 from this procedure because it did not exhibit {\pab} emission at all epochs). We then averaged the offsets measured for GY~292 and GY~308 at each epoch, and shifted all spectra by that amount to center {\pab} at its rest wavelength. The typical shift applied was about 2~{\AA}, or roughly half a pixel, comparable to the uncertainty in the measured {\pab} line centers. The average difference between the offsets derived for GY~292 and GY~308 was 1.7~{\AA}, with no systematic behavior in the size or direction of observed shifts. {\pab} is separated from the He~$\lambda 10830$ line by a few hundred pixels in our spectra, so we also measured the offset of the {\pag} feature from its rest wavelength at each epoch in order to test the appropriateness of this wavelength recalibration ({\pag} is much closer to the \ion{He}{1} line, but is considerably weaker than {\pab}, providing far less accurate wavelength centroids). The difference between the {\pab} and {\pag} offsets was on average about $1.4$~{\AA} -- less than the uncertainty on the {\pab} centroids -- demonstrating that these wavelength corrections are robust at the sub-pixel scale in the wavelength regime of \ion{He}{1}.

Having verified the fidelity of the wavelength solution near the \ion{He}{1}~$\lambda 10830$ line, we characterized the structure of the \ion{He}{1} line profile in each of our target spectra by measuring the EW of the redshifted and blueshifted portions of the line separately. The wavelength regions adopted to sample the red and blueshifted portions of the line profile are listed in Table \ref{tab:lineparams}, and shown in Figure \ref{fig:HeI}, where we present close-ups of the time evolution of each star's \ion{He}{1}~$\lambda 10830$ and {\pab} line profiles. We selected these wavelength regions so as to best capture the change in morphology of the \ion{He}{1} line on each side of the rest velocity while avoiding nearby contaminating features.

\subsection{SED analysis}
\label{sec:diskemission}

To quantify the relative contributions that the stellar photosphere and circumstellar disk make to the [3.6] light curve, we have constructed and interpreted broadband SEDs for GY 292, 308, and 314 by combining our \textit{Spitzer}/IRAC data with optical and NIR photometry reported in the literature. $R$ \& $I$~band photometry is taken from Gordon \& Strom (1990, unpublished; but reported in Table~2 of \citealp{Wilking05}), $JHK_s$ photometry is taken from the 2MASS database \citep{Skrutskie06}, and median IRAC magnitudes are adopted from the light curves reported here. 

\begin{figure}
\includegraphics[angle=90,width=\linewidth]{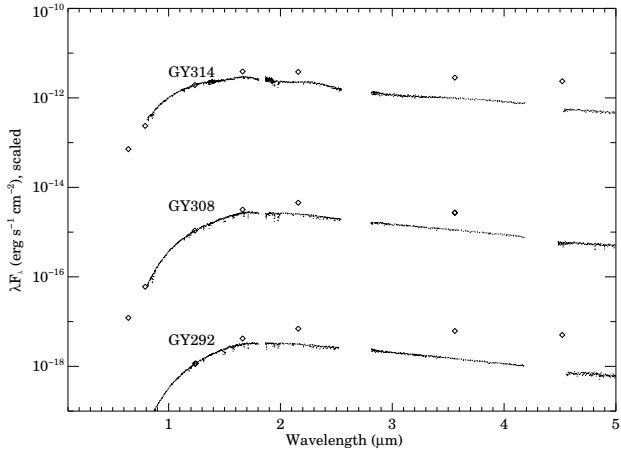}
\caption{\small{Optical/IR SEDs for GY 292, 308 and 314 (diamonds), including $RIJHK_s$ from the literature and our \textit{Spitzer} IRAC photometry. Each SED is compared to the 0.8--5.0 $\mu$m spectrum of a diskless field star of the same spectral type, artifically reddened with the \citealp{Fitzpatrick99} extinction law (with $R_V=3.1$) to match the target star's spectroscopically inferred A$_H$, and normalized to the match the target star's $J$~band flux. GY 292, 308, and 314 all possess near- and mid-infrared excesses relative to their reddened spectral standards, implying that about 76\%, 59\%, and 64\% of their [3.6] emission, respectively, arises from their circumstellar disks.}}
\label{fig:photflux}
\end{figure}

The resulting SEDs for GY 292, 308, and 314 are shown in Figure~\ref{fig:photflux}, where we compare each YSO SED to a 0.8--5.0 $\mu$m diskless spectral standard of the appropriate spectral type from the IRTF spectral library \citep{CRV05}. Each standard spectrum was artificially reddened to simulate the extinction we measure toward each YSO (see \S \ref{sec:spt} and Table~\ref{tab:mdot}) using the \cite{Fitzpatrick99} extinction law with an assumed $R_V = 3.1$ (though the result is largely insensitive to the adopted $R_V$ value; see further discussion later in this section). We then computed integrated fluxes at $JHK_s$ and IRAC [3.6] and [4.5] by convolving the spectra with the appropriate filter response curves. The standard spectrum was normalized so as to match the YSO flux at $1.24~\mu$m, where the contribution due to veiling is low.

Assuming the spectral standards serve as reasonable proxies for the photospheric components of our YSO fluxes, the emission arising from the circumstellar disk at $3.6~\mu$m is then $F_{3.6}-F_{\mathrm{phot},3.6}$, where $F_{3.6}$ is the object's median [3.6] flux and $F_{\mathrm{phot},3.6}$ is the integrated flux of the reddened spectral standard in the IRAC [3.6] band. This modeling predicts that the disk contributes 76\% (GY 292), 59\% (GY 308) and 65\% (GY 314) of each YSO's total $3.6~\mu$m flux. The circumstellar disk is even more dominant at IRAC [4.5], where a similar calculation implies disk fractions of 91\% (GY~292) and 85\% (GY~314; note that we did not obtain [4.5] data for GY~308). We also estimated the expected amount of veiling ($r_\lambda \equiv F_{\mathrm{disk},\lambda}/F_{\mathrm{phot},\lambda}$) in the $H$ and $K_s$ bands by comparing the filter-integrated fluxes of the reddened spectral standards to the YSO 2MASS fluxes, deriving \{$r_H$,$r_{K_s}$\} veilings of \{0.4,1.3\}, \{0.2,0.8\}, and \{0.4,0.7\} for GY 292, GY 308, and GY 314, respectively. These values should be considered lower limits, as they assume negligible veiling at $J$~band (discussed further later in this section), and they are consistent (as lower limits) with the more robust results we derived through direct measurements of absorption lines in \S \ref{sec:VeilVar}.

These SED-based veiling/disk emission estimates are subject to three dominant sources of uncertainty: the adopted extinction law, potential presence of $J$~band veiling, and long-term photometric variability. However, as we demonstrate below, these effects do not appear large enough to alter our basic conclusion that, for each each YSO, the mid-IR flux is dominated by disk emission:

\begin{itemize}
\item{\textit{Adopted extinction law:} Our SED analysis depends centrally on the artificially reddened spectral standard that serves as a proxy for flux expected from the YSO's stellar photosphere. By directly affecting the shape of that assumed photospheric flux, errors in either the magnitude or wavelength dependence of the assumed extinction will translate directly into errors in our inferred veilings or disk emission fractions. Several previous studies, however, have established that the NIR extinction law is highly insensitive to dust properties, or equivalently to the choice of $R_V$ \citep[e.g.,][]{Fitzpatrick99,Indebetouw05,Chapman09}. We confirmed this result by re-calculating disk fraction and veiling estimates after reddening the standard with an $R_V=5.5$ (i.e., grayer) extinction law, and the resulting [3.6] disk fractions decreased by only 1--2\%, while veiling estimates decreased by no more than $\Delta r_{\lambda}=0.1$.}

\item{\textit{Potential $J$~band veiling:} Our SED analysis implicitly assumes that each YSOÕs $J$~band flux is entirely photospheric, i.e., $r_J=0$. The estimate we infer for the fraction of each YSO's flux that is non-photospheric in nature is therefore quite conservative: if a YSO does indeed have a nonzero $J$~band veiling \citep[as most do: many CTTSs possess $r_J \sim$0.1--0.2;][]{Cieza05}, the disk excesses inferred from our SED based analysis will be underestimated by a factor proportional to the unacknowledged $J$~band excess. Specifically, to correct our estimates for a YSO with a typical $J$~band veiling of $0.2$ (corresponding to a flux correction of about 15--20\%), we would need to increase $r_H$ by 0.2--0.3, $r_K$ by 0.3--0.5, and the [3.6] disk fraction estimate by 4--8\%. Thus, the results we derive above should be considered robust lower limits on disk fraction and veiling, and the mid-IR emission from the circumstellar disk is likely to be \textit{even more dominant} at 3.6~$\mu$m than we predict in our sample of YSOs.}

\item{\textit{Long term photometric variability:} The accuracy of the broadband SEDs we have constructed is limited by the amplitude of each YSO's long-term photometric variability, due to the approximately ten-year timespan between the 2MASS $JHK_s$ and our \textit{Spitzer} IRAC measurements. Such long-term variations are indeed observed, though not ubiquitous, in young stars; for example, \cite{Parks_prep} find that 31\% of the variable stars in their study of $\rho$~Oph exhibit modest to significant $K_s$~band variations over timescales of a few months to 2.5 years, although the long-term variability amplitude is limited to $\lesssim 0.3$~mag in the majority of cases. Furthermore, \cite{Scholz12} find that there are very few YSOs (2--3\% of cluster members in several nearby star-forming regions including $\rho$~Oph) that are highly variable ($>0.5$~mag) over years-long timescales. To test the potential effect of long-term variability in our sample, we compared the 2MASS magnitudes of our objects (listed in Table~\ref{tab:basicdata}), which were measured between 1997--2001, to the photometry obtained in 1993--1994 by \cite{Barsony97}, and find changes in all three NIR bands of only about $0.1$~mag on average. \cite{Scholz12} predict that the absolute error in a particular photometric measurement of a YSO due to long-term variability is 5--20\%, with some dependence on age and the fraction of emission produced by the disk. As the discussion of potential uncertainties due to $J$~band veiling demonstrates, errors on the 15--20\% level limit the precision of the exact disk fraction values derived above, but the qualitative result that the majority of [3.6] emission originates in the circumstellar disk is robust against potential long-term photometric variations.}
\end{itemize}

\section{Discussion}
\label{sec:discussion}

\begin{figure}
\includegraphics[width=\linewidth]{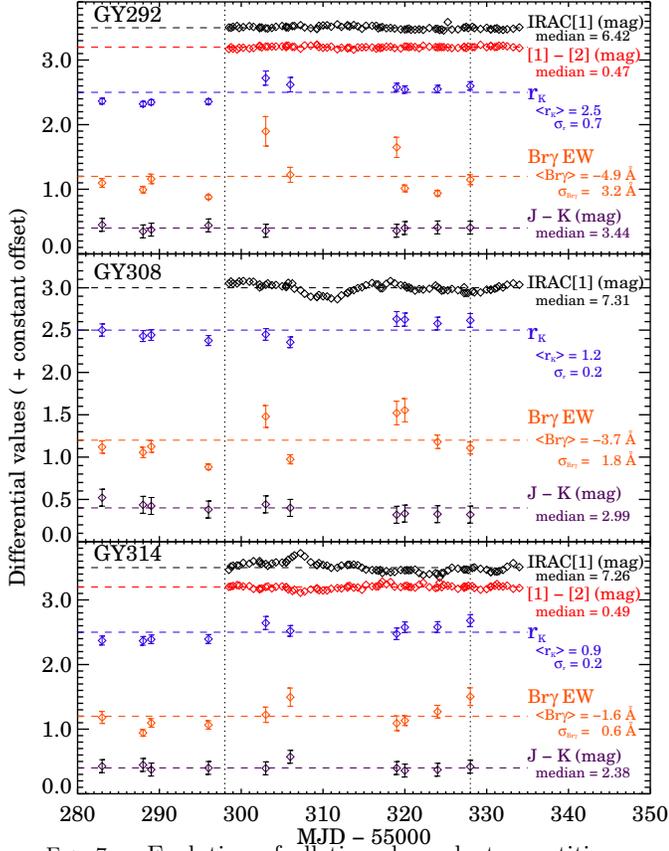}
\caption{\small{Evolution of all time-dependent quantities measured from the light curves and spectra of GY 292, 308, and 314: IRAC 1 (i.e. [3.6]) photometry (black), IRAC ($1-2$) color (red), $K$~band veiling (blue), {\brg} EW (accretion rate diagnostic, orange), and spectrally integrated $J-K_s$ color (extinction diagnostic, purple). The average value is listed to the right of each time series, and indicated on the plot by a dashed line; the plot displays fractional deviations from this average. The magnitude of variations in $r_K$ and {\brg} EW have been suppressed by a factor of two for clarity. GY 308 and 314 exhibit significant mid-IR variability, and GY 292 and 308 exhibit variability in the accretion-sensitive Br $\gamma$ emission lines; we identify no significant correlations between the time dependence of these physical diagnostics, however. }}
\label{fig:master}
\end{figure}

We now seek to use the time-averaged and time-resolved spectral diagnostics measured in the previous section (e.g, $H$ \& $K_s$~band veiling, {\brg}, {\pab}, and \ion{He}{1}~$\lambda10830$ EWs, and spectrally integrated $J-K_s$ colors) to inform our understanding of GY 292, 308, and 314's mid-IR light curves and evolutionary states. After a brief discussion of the implications of the time-averaged properties of our YSOs, we test for signatures of the physical mechanisms driving the moderate mid-IR variability in our sample by computing the linear Pearson's correlation coefficient between their IRAC light curves and the relevant spectral diagnostics. The time dependence of our full suite of spectral and photometric measurements is summarized for these three YSOs in Figure \ref{fig:master}.

\subsection{Evolutionary State}
\label{sec:ages}

Theory and observations suggest that YSOs begin their lives as strong accretors with massive disks, with the accretion rate and disk mass declining over time as circumstellar material is depleted by accretion, dispersed by stellar outflows, or collected into dense planetesimals. All three of GY~292, GY~308, and GY~314 were classified as Class~II sources (i.e. T~Tauri stars) by \cite{Bontemps01}; although we lack the longer wavelength mid-IR coverage to reliably compute $\alpha$, the spectral index commonly used to classify YSOs, the SEDs constructed and disk fractions computed in \S \ref{sec:diskemission} are qualitatively consistent with this classification (see also Figure \ref{fig:photflux}). Star formation is a dynamic process, with notable examples of young stars whose {\Macc} and veiling have undergone significant changes~\citep[e.g.][]{GL96,CG10}. We therefore define a young star's ``evolutionary state'' as a measure of the source's characteristic levels of veiling and mass accretion, such that heavy accretors/veiled young stars are less ``evolved'' than their more quiescent kin. This evolutionary sequence will not, in general, map well to a star's chronological age, but instead to its present day levels of disk accretion and emission. The time-averaged veiling and mass accretion rates of our three YSOs differ measurably, such that our data suggest a clear ``evolutionary sequence'' that coincidentally follows their numerical designation: GY~292 has the highest accretion rate and veiling, GY~314 has the lowest accretion rate and veiling, and GY~308 is intermediate between the two.

Veiling has been shown to correlate strongly with the evolutionary state of a forming star: the most embedded objects are shrouded in so much circumstellar material that photospheric features can disappear entirely, while disk emission decreases gradually as matter is accreted onto the evolving star~\citep[e.g.,][]{GL96}. Our veiling measurements displayed graphically in Figure~\ref{fig:allveil} agree well in both scale and spectral shape with the disk excesses implied by the SED analysis in \S \ref{sec:diskemission}, and identify GY~292 as the most heavily veiled (and thus the YSO having the most dominant circumstellar disk) of these three; GY 314 is the least veiled, but only modestly less so than GY~308.

The evolutionary sequence suggested by the level of veiling in our YSO targets is also consistent with the mass accretion rates measured in Section \ref{sec:EmLineVar} and shown in Table \ref{tab:mdot}. GY~292 has the highest time-averaged mass accretion rate (10$^{-8.5}$ M$_{\odot}$ yr$^{-1}$), and GY~314 has the lowest (10$^{-9.2}$ M$_{\odot}$ yr$^{-1}$). Additionally, the amplitude of variability in {\Macc} (as inferred from the standard deviation of the EW({\brg}) time series, $\sigma_{\mathrm{Br}_{\gamma}}$ in Figure~\ref{fig:master}) is largest in GY~292 and smallest in GY~314, hinting that higher characteristic accretion rates may also be accompanied by more pronounced short-timescale {\Macc} variations. We do note, however, that while GY~292's near-infrared spectrum exhibits notable variability, particularly in the \ion{H}{1} and \ion{He}{1} accretion diagnostics, it is relatively quiescent in both IRAC bands over the \textit{Spitzer} observation window. This suggests that, despite the broad evolutionary trends identified by \cite{MoralesCalderon11}, mid-IR variability is not a uniform, monotonic function of accretion or disk activity, at least on week- to month- timescales.

The mass accretion rate (as inferred from {\brg} EW) of each YSO in our sample varies by a factor of 3--5 over our observation window, while time-averaged accretion rates differ by about a factor of five across the sample. Since the latter are averaged quantities, this suggests a real difference in evolutionary state of the three YSOs that is more significant than the short-term, stochastic variations observed in any one YSO. While this is not strict proof of evolutionary ordering -- indeed, Class~II YSOs have been seen to exhibit differences in EW({\brg}) and EW({\pab}) of up to a factor of $\sim$10 across a flux-limited sample \citep{GL96} -- the fact that our objects seem to be measurably more disparate in their characteristic accretion activity than any single YSO is variable, combined with a clear difference in the level of veiling across the sample, leads us to propose the evolutionary sequence above as the simplest (but not only) explanation that is consistent with our measurements.

\subsection{Possible explanations for mid-IR variability}

In the following sections, we explore if each YSO's observed IRAC [3.6] variability can be attributed to physical processes that correlate with the its veiling, extinction, and/or accretion rates.

\subsubsection{Veiling correlation}
\label{sec:veilcor}

Detailed radiative transfer models indicate that a YSO's NIR and mid-IR excess emission both likely arise from the innermost regions of its circumstellar disk. \cite{Espaillat09} showed that more than 90\% of the $7~\mu$m flux produced by a typical circumstellar disk originates inward of $\sim$0.25~AU; excess emission at shorter wavelengths will be even more centrally condensed, arising from the warmest material at the smallest radii. As disks within systems with {\Macc} of $10^{-9}$ to $10^{-8} \Msun/$yr are typically truncated at an inner radius of 0.08 to 0.15~AU \citep[e.g.,][]{Espaillat07,LCS11}, this implies that the vast majority of both the $K$~band and $3.6~\mu$m flux is emitted within the inner $\sim$0.05--0.20~AU of the disk, likely from the inner wall at the disk's truncation radius. Thus one might naturally expect [3.6] magnitude and NIR veiling to be correlated.

To test this possibility in our sample, we explicitly look for a correlation in the time evolution of each YSO's [3.6] flux and NIR veiling $r_H$ and $r_K$. Interpolating the IRAC [3.6] light curve onto the spectroscopic epochs, we calculate the linear Pearson's correlation coefficient, $r$, between the two time series; we find $r=0.59$ and $0.38$ (GY 292), $0.30$ and $0.21$ (GY 308), and $0.31$ and $0.23$ (GY 314), where the two $r$values listed for each object are for $r_H$ vs. [3.6] and $r_K$ vs. [3.6], respectively. These values are well below the 95\% confidence value of $|r| = 0.811$, and thus we are unable to confidently link the observed mid-IR variability with changes in $H$ and $K$ band veiling. However, veiling variability is only present at a modest level in all three objects (in contrast with accretion variability, which is clearly evident in at least two YSOs; see \S \ref{sec:acccor}). Thus modest changes in disk emission could still potentially explain the IRAC light curves of marginally variable objects such as GY~292.

\subsubsection{Extinction model}
\label{sec:extinct}

We next investigate whether changes in each YSO's line-of-sight extinction could explain the observed mid-IR variability. To explore this possibility, we construct a simple extinction model with three assumptions: (1) the median IRAC [3.6] magnitude represents the source's median state; i.e., the spectral parameters inferred earlier and published 2MASS magnitudes are a good representation of that epoch; (2) the variability in the measured IRAC [3.6] magnitude is due solely to changes in extinction; and (3) the interstellar extinction law derived by \cite{Indebetouw05} is an appropriate proxy for disk extinction\footnote[5]{The \cite{Indebetouw05} and \cite{Fitzpatrick99} extinction laws agree to within a few percent in the infrared, thus we adopt the former for this model since it is native to the IRAC filter set.}. We parameterize extinction as $J-K_s$ color, and compare the model predictions to the spectrally integrated $J-K_s$ colors computed in \S \ref{sec:spt}.

\begin{figure}
\includegraphics[width=\linewidth]{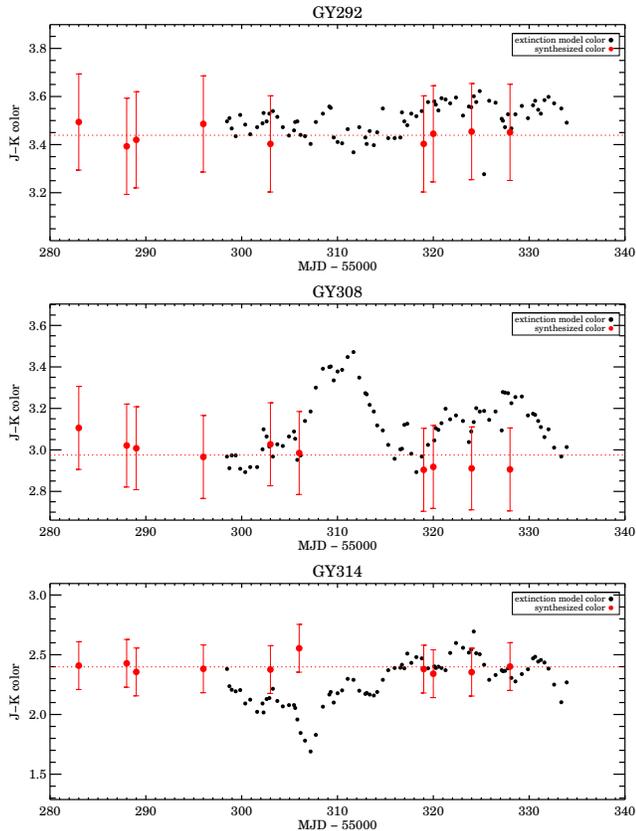}
\caption{\small{Evolution of $J-K_s$ color (derived by integrating the SpeX spectra over the 2MASS filter bands; large red circles) superposed on the variations predicted for the $J-K_s$ color from a simple model in which variations in each YSO's [3.6] light curve are solely attributed to changes in the YSO's extinction, $A_{[3.6]}$ (small black circles). Extinction variations could induce GY 292's low level of [3.6] variability, but attributing GY 314's observed [3.6] variations to variable extinction predicts variations in $J-K_s$ colors that are inconsistent with the observed SpeX data. Modeling GY 308's [3.6] varations as extinction events predicts $J-K_s$ variations that are only marginally consistent with the observed SpeX spectral data.}}
\label{fig:extinct}
\end{figure}

In particular, we used our [3.6] measurements with the above assumptions to calculate the extinction-based model $A_J$ and $A_{K_s}$ at each IRAC epoch, then computed the $J-K_s$ color predicted by the model as $(J-K_s)_{\mathrm{model}} = (J-K_s)_{\mathrm{2MASS}}+(A_J - A_{K_s})$. We compare these model colors to the spectrally integrated $J-K_s$ colors computed in \S \ref{sec:spt} in Figure~\ref{fig:extinct}. As the figure demonstrates, the spectrally integrated $J-K_s$ colors show no significant variability across the spectroscopic epochs, limiting extinction-driven changes in $J-K_s$ color to less than $\sim$0.2~mag for all three YSOs. Modeling the [3.6] light curves as fully due to extinction variations, by contrast, predicts $J-K_s$ variations at the 0.2--0.8~mag level over the full duration of the photometric monitoring. The $J-K_s$ variations predicted by the extinction model are definitively inconsistent with GY 314's spectrally integrated $J-K_s$ colors, ruling out extinction as a potential mechanism to explain the structure observed in GY~314's [3.6] light curve. GY~292 is not significantly variable at [3.6], by contrast, constraining any potential extinction variations to $\lesssim 0.2$~mag in $J-K_s$ for this YSO. GY~308 is an intermediate case; while the model and spectrally integrated $J-K_s$ are largely consistent at most of the spectral epochs, there are spectroscopic datapoints that disagree with the predictions of the extinction-based model at the 1.5 $\sigma$ level, suggesting that mechanisms other than extinction must be at least partially responsible for the observed mid-IR variability.

The weakest link in the above calculation is the assumption that an interstellar extinction law accurately replicates the extinction due to dust in a circumstellar disk. Extinction curves for circumstellar dust are often thought to be grayer than for interstellar dust \citep[e.g.,][]{Flaherty07}, so a model based on an interstellar extinction law could potentially overestimate the scale of the expected NIR extinction variations required to reproduce a given [3.6] light curve due to extinction from intermediate disk structures. Nevertheless, the exact wavelength dependence of extinction in disks is not well-constrained and dust properties could vary significantly with evolutionary age, stellar properties, or viewing angle~\citep[e.g.,][]{Sung09}, so the above calculation serves as a well-defined test of an extreme case of disk extinction as the source of mid-IR variability.

\subsubsection{Accretion rate correlation}
\label{sec:acccor}

Enhanced accretion activity could potentially heat the inner disk, both through viscous dissipation and via energetic radiation emitted by the accretion shock, and thereby produce an increase in the mid-IR emission from the disk's inner edge. Physical changes in the disk structure related to the accretion process could also alter the disk's emission properties. If active, these mechanisms should produce a correlation between the YSO's [3.6] magnitude and the strength of accretion-sensitive lines, such as {\brg}.

As discussed in \S \ref{sec:ages}, the mass accretion rate, as traced by {\brg} EW, appears to vary significantly with time in our sample, especially in GY 292 and GY 308. To formally quantify that EW({\brg}) is variable in our YSOs, we adopt the null hypothesis that {\brg} EW is constant in time, and perform a $\chi^2$ minimization to fit our measured {\brg} EWs with a constant accretion rate model. This model produces reduced $\chi^2$ values of 29, 28, and 3.7 for GY~292, 308, and 314, respectively. The significant deviation from the time-independent accretion model is strong evidence that these YSOs undergo measurable changes in {\Macc} on day- and week- timescales, particularly the presumably less evolved YSOs GY~292 and 308. 

To test if the variability in the accretion rate is related to the structure in the [3.6] light curve, we calculate the level of correlation between each YSO's observed {\brg} EW and [3.6] flux, following the procedure of \S \ref{sec:veilcor} above. We calculate $r$values of $r = -0.31$ (GY 292), $-0.59$ (GY 308), and $-0.41$ (GY 314), each well below the threshold values $|r| = 0.811$ (6 data points) or 0.878 (5 data points) that indicate 95\% confidence in a correlation. We are thus unable to establish that the observed changes in mass accretion are primary contributors to each YSO's mid-IR variability.

\begin{figure*}$
\begin{array}{ccc}
\includegraphics[width=0.33\linewidth]{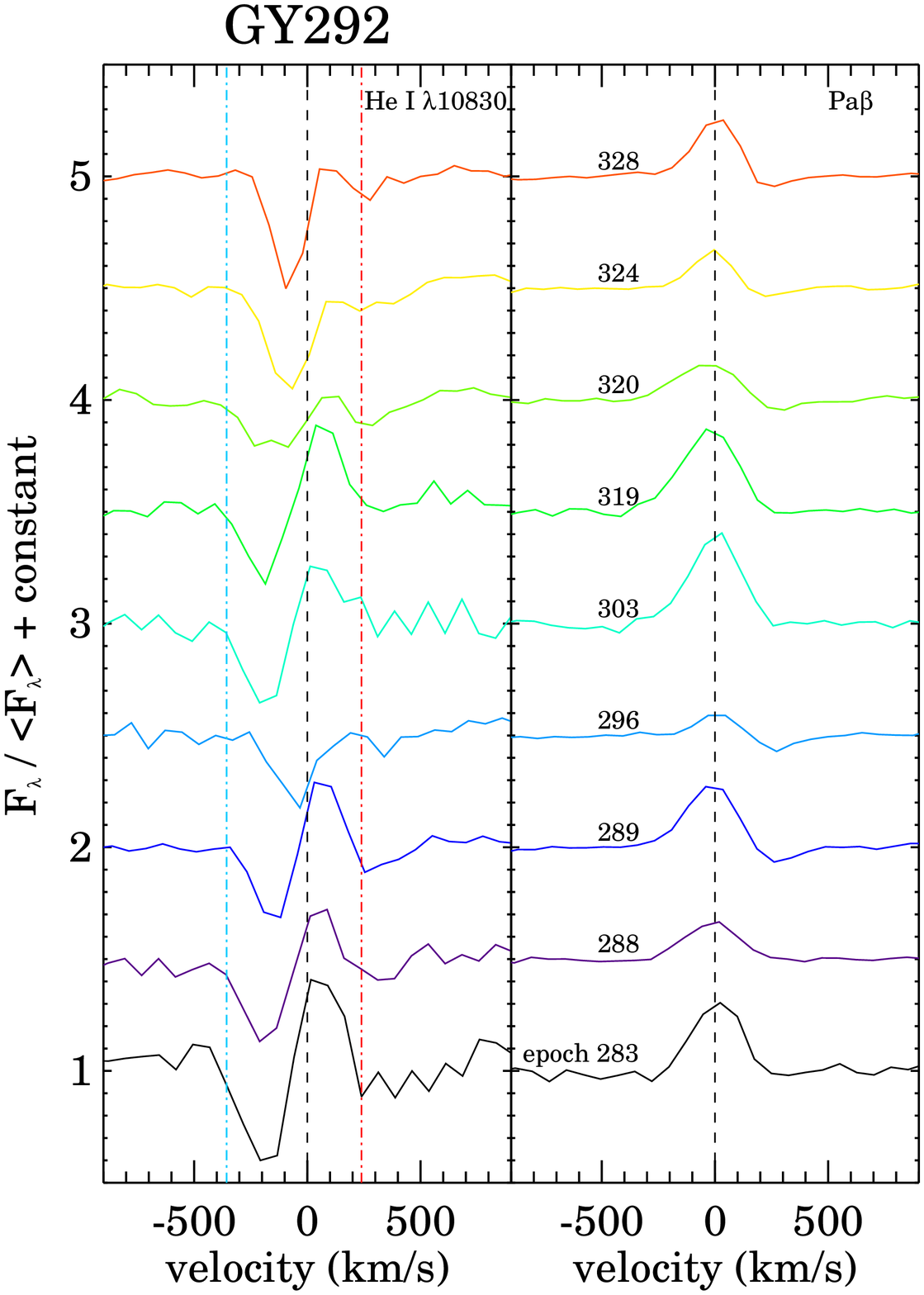}&
\includegraphics[width=0.33\linewidth]{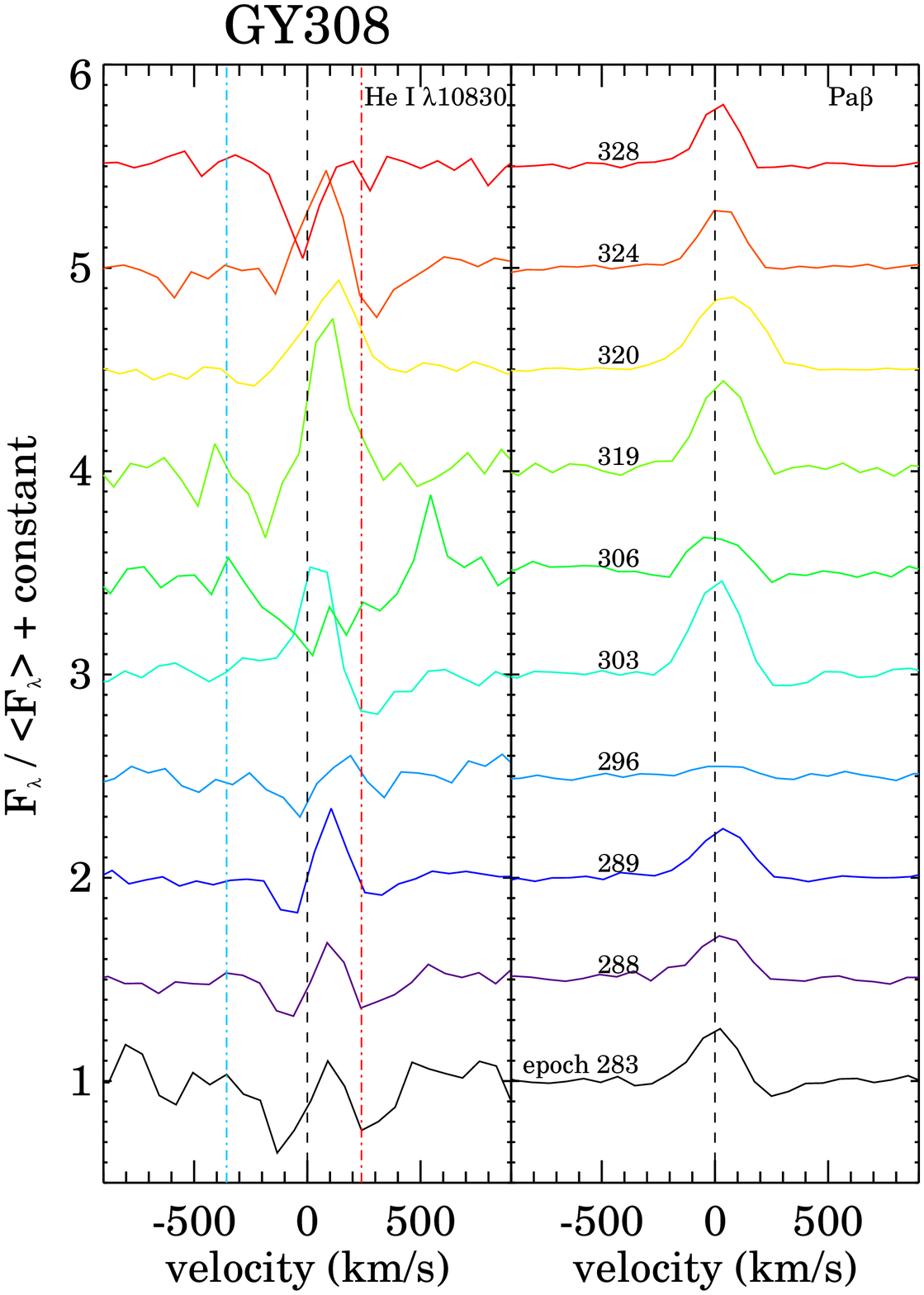}&
\includegraphics[width=0.33\linewidth]{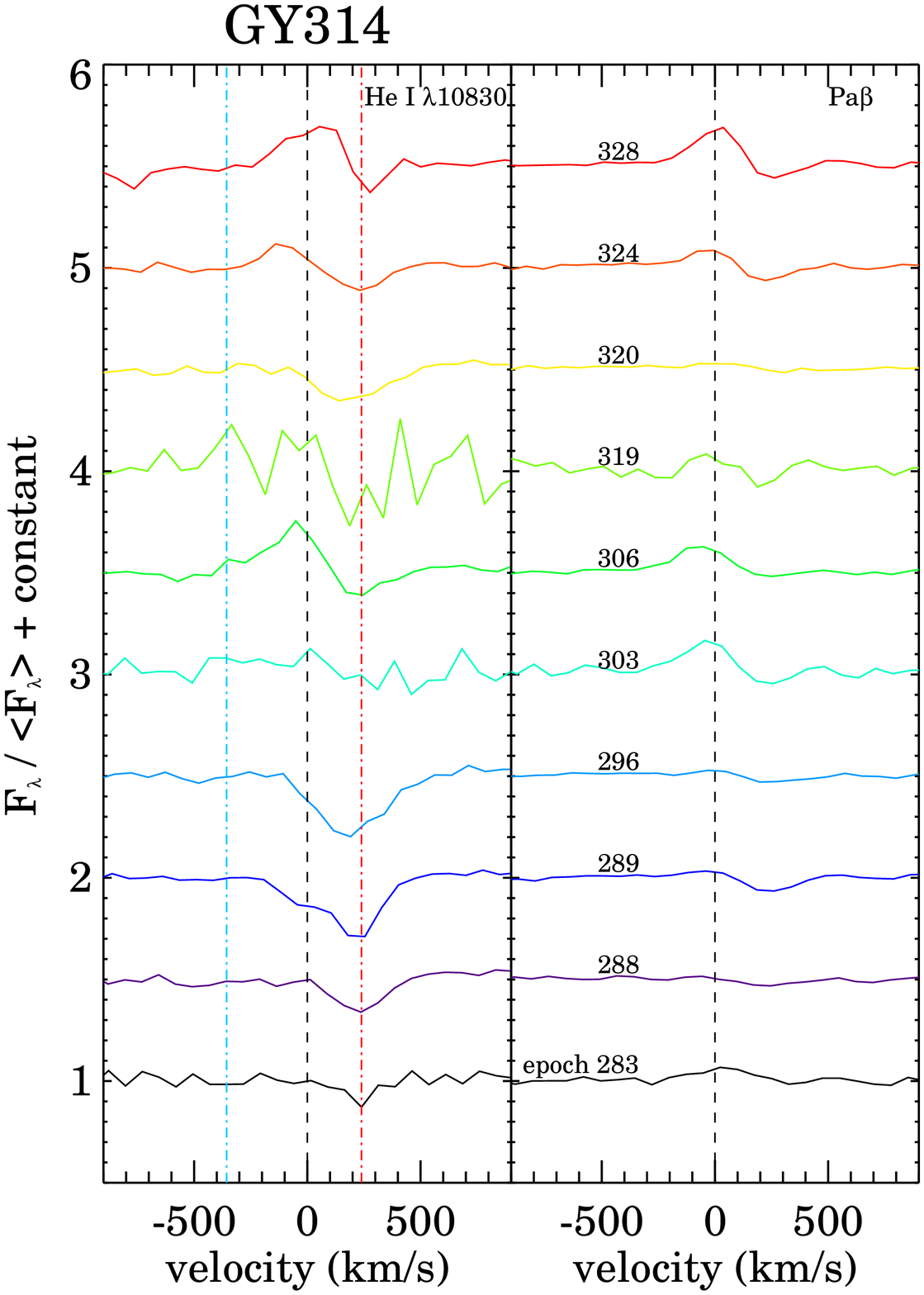}
\end{array}$
\caption{\small{ \ion{He}{1} $\lambda 10830$ and {\pab} profiles for each of our targets. The blueward (redward) \ion{He}{1} EW was computed over the spectral region between the dotted blue (red) line and the dotted black line at the systemic rest velocity. All spectra were shifted so as to center the {\pab} line center at its rest wavelength. Redshifted \ion{He}{1} emission appears to be correlated with {\pab}, while blueshifted absorption is less so (see also Figure~\ref{fig:hecorr}). The epoch is MJD$-55000$.}}
\label{fig:HeI}
\end{figure*}

\subsection{\ion{He}{1} $\lambda 10830$}
\label{sec:HeI}

We conclude this study by examining the temporal correlation between the {\pab} emission line and the structure of the \ion{He}{1}~$\lambda10830$ line. Previous observational and theoretical work has shown that YSO \ion{He}{1} and \ion{H}{1} line profiles can both be shaped by accreting and outflowing material \citep[e.g.,][]{Alencar05,KEF07,KRH11}; nonetheless, the typical emergent profiles seen in these lines from YSOs differ significantly, as do their assumed sensitivity to accretion and outflow.

{\pab} is seen in emission in most Class~II YSOs, with a minority exhibiting a modest amount of sub-continuum redshifted absorption (e.g., \citealp{FE01} found that 42/50 CTTS exhibited {\pab} in emission, of which 13 also exhibited redshifted absorption). {\pab} line strengths also correlate quite well with mass accretion rates inferred from blue continuum excesses \citep[e.g., Fig.~3,][]{MHC98}, such that it is often adopted as an accretion indicator, despite the fact that much of the observed emission could plausibly arise from outflowing material \citep[see, e.g., Fig.~9,][]{KR12}.  

\ion{He}{1} $\lambda10830$, in contrast to {\pab}, exhibits absorption components just as frequently as emission: in their analysis of high resolution \ion{He}{1} $\lambda10830$ profiles of 39 CTTSs, \cite{Edwards06} found 35 and 33, respectively, that exhibited absorption and emission components. Studies of \ion{He}{1} $\lambda10830$'s excitation conditions indicate that outflows are likely responsible for most of the emission, with outflows and accretion responsible for blueshifted and redshifted absorption, respectively \citep{Fischer08,KF11}. It has been suggested, however, that the outflows traced by \ion{He}{1} may themselves be accretion-driven \citep[e.g.,][]{Edwards06,MP08}, suggesting that correlations may exist between the (presumably) accretion-sensitive \ion{H}{1} lines, such as {\pab}, and the predominantly outflow-sensitive lines, such as \ion{He} {1}~$\lambda10830$.  

\begin{figure*}$
\begin{array}{ccc}
\includegraphics[width=0.18\linewidth,angle=90]{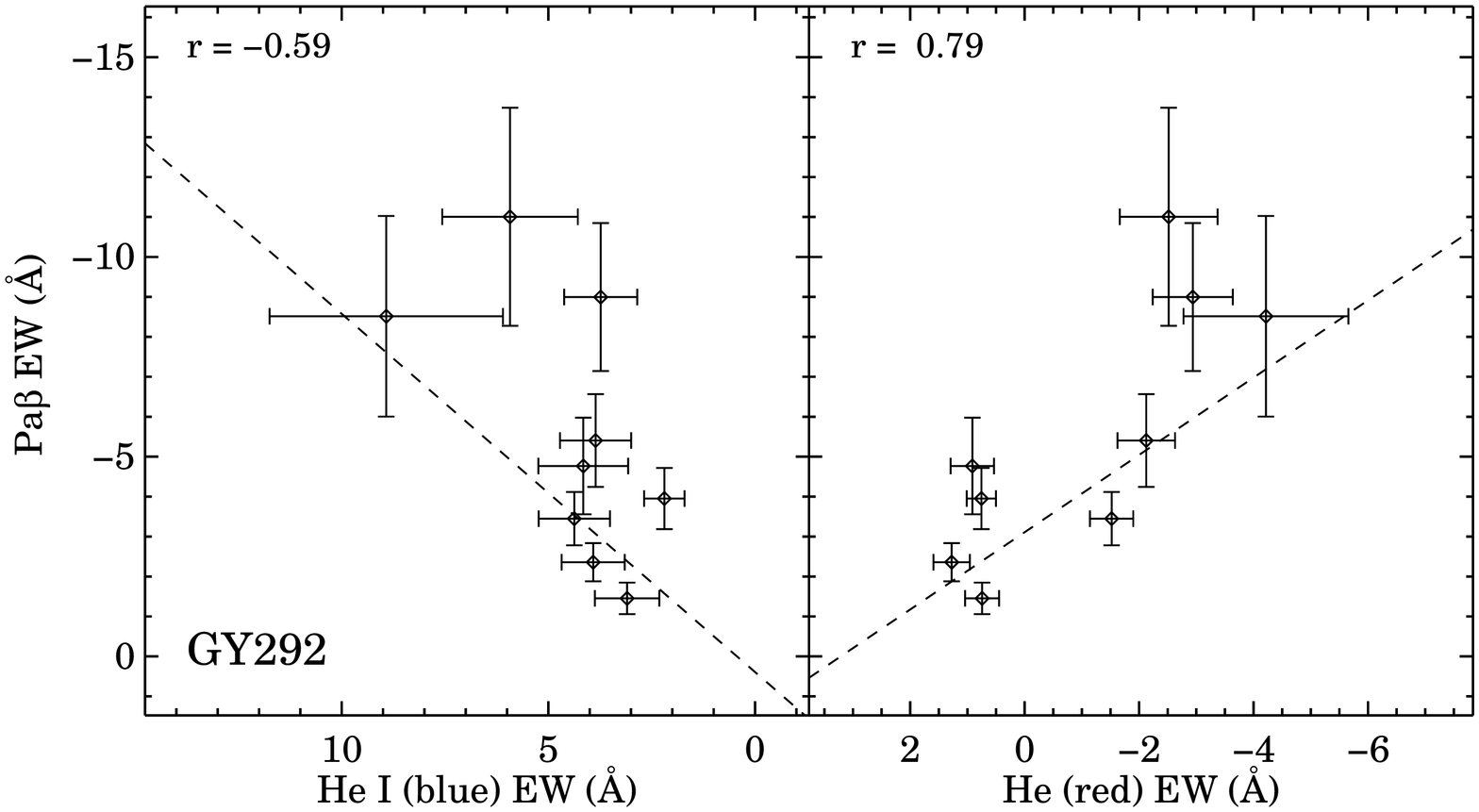}&
\includegraphics[width=0.18\linewidth,angle=90]{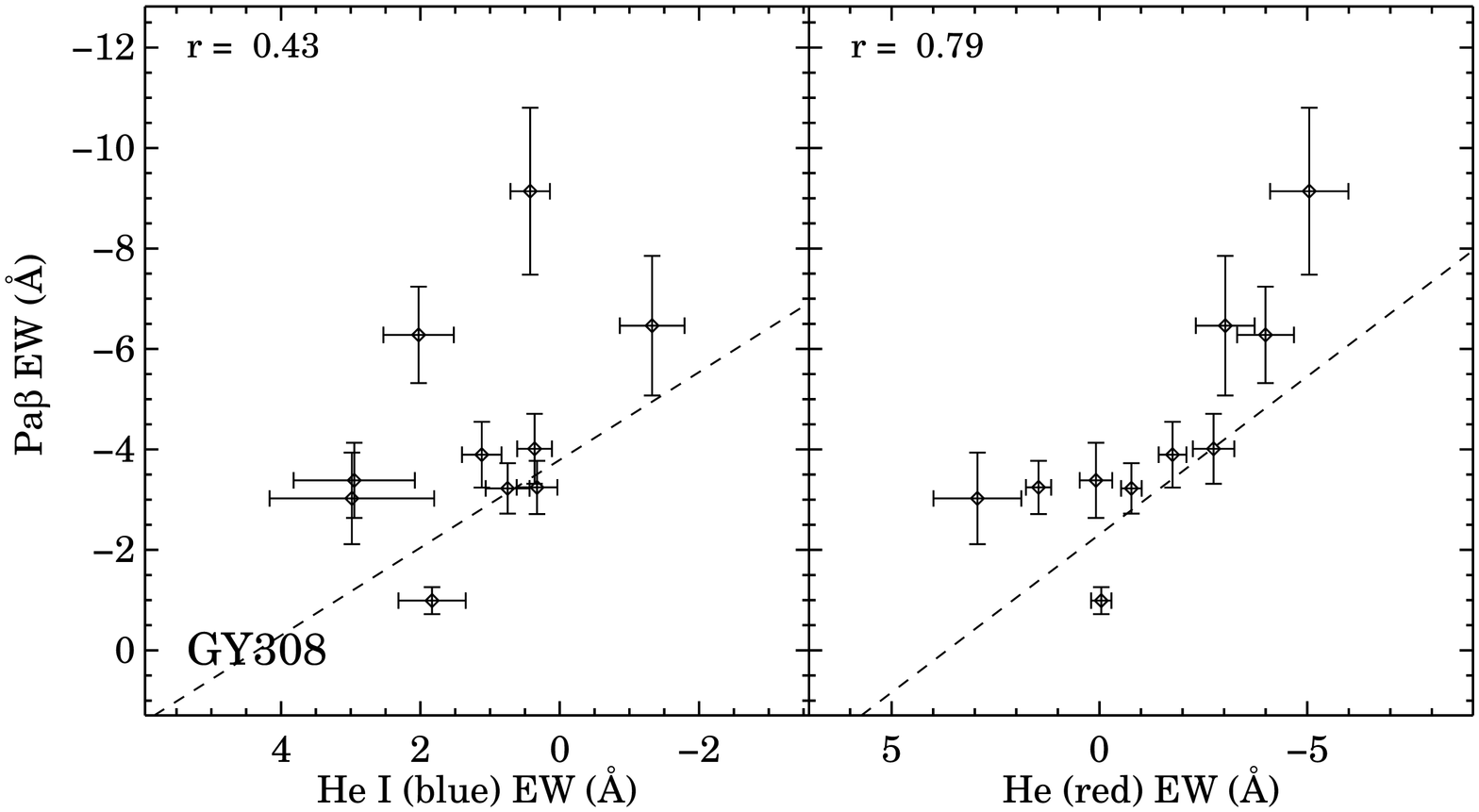}&
\includegraphics[width=0.18\linewidth,angle=90]{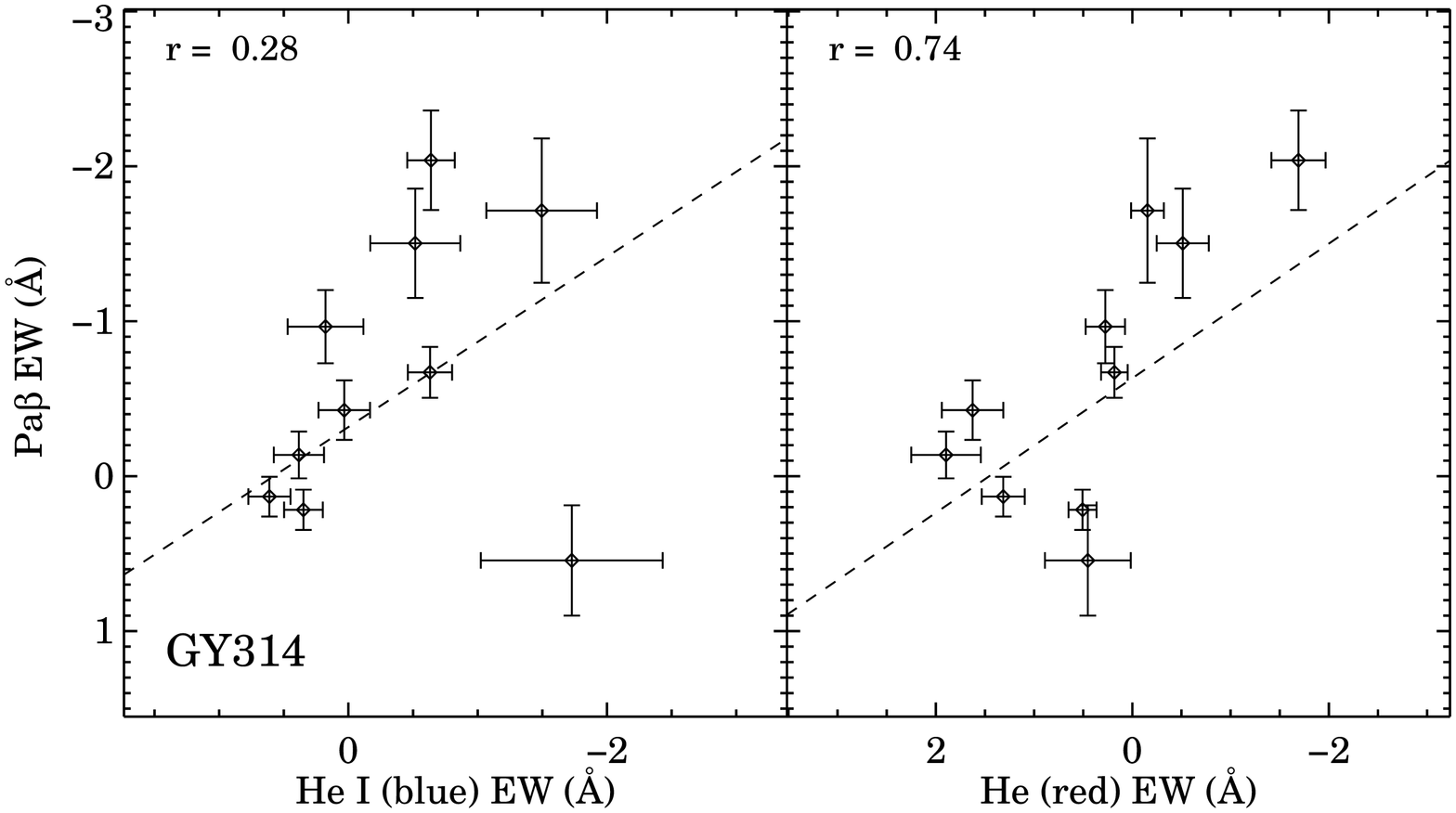}
\end{array}$
\caption{\small{Linear fits to the relations between {\pab} equivalent width (an accretion diagnostic) and each of \ion{He}{1}~$\lambda 10830$ blueshifted absorption (left panels) and \ion{He}{1}~$\lambda 10830$ redshifted emission (right panels) for the YSOs GY 292, GY 308, and GY 314. Correlation coefficients $r$ with magnitude greater than about 0.6 indicate statistically significant correlations, suggesting that the redshifted portion of \ion{He}{1} may be more closely correlated with accretion activity in these YSOs than the blueshifted portion. Note that negative EWs indicate emission, while positive EWs measure absorption.}}
\label{fig:hecorr}
\end{figure*}

To test for potential correlations between the kinematically simple {\pab} emission line, and the kinematically richer \ion{He} {1}~$\lambda10830$ line, we calculated linear Pearson's correlation coefficients ($r$) between the EW of the full {\pab} line and the EWs of the blueshifted and redshifted components of \ion{He}{1} (computed in the regions bounded by the black dotted line, which is at the systemic rest velocity, and the blue and red dotted lines, respectively, in Figure~\ref{fig:HeI}). These constant-width windows were selected such that the entire P-Cygni profile, where present, was captured at all epochs. {\pab} and \ion{He}{1} EWs were corrected for the effects of veiling using the procedure of \S \ref{sec:VeilVar} and measurements of the \ion{Al}{1}~$\lambda\lambda 13123,13151$ doublet in both our target and standard star spectra. The results of these correlation calculations are shown visually in Figure~\ref{fig:hecorr}.

We find a clear correlation between the {\pab} EW and the EW of the red side of \ion{He} {1}~$\lambda10830$ for all three YSOs ($r = 0.79$, 0.79, and 0.74 for GY~292, 308, and 314, respectively), but only a modest (and not statistically significant) correlation between the blueshifted side of this \ion{He}{1} line and {\pab} emission ($r=-0.59$, 0.43, and 0.28; $|r| \gtrsim 0.6$ indicates statistical significance for this sample size). For all three YSOs, {\pab} emission increases more or less monotonically with the strength of the redshifted component of the \ion{He} {1}~$\lambda10830$ line, suggesting that both of these line components are indeed sensitive to accretion activity, consistent with the interpretation advanced by \cite{Fischer08}. This could reflect either a) the presence of increased \ion{He} {1}~$\lambda10830$ emission from winds at epochs with stronger {\pab} emission, which itself could be due to a combination of emission from accretion or outflows, or b) a geometrical effect whereby emission from {\pab} is enhanced during epochs when the magnetospheric accretion column is perpendicular to our line of sight, and therefore is also unable to produce significant redshifted absorption in \ion{He} {1}~$\lambda10830$.  

We note, however, that as each of our targets is a ``typical'' Class~II YSO, the correlation we detect between {\pab} emission and redshifted \ion{He} {1}~$\lambda10830$ may only hold for sources over a limited range of evolutionary states, or with a particular source geometry. Indeed, FU Ori stars demonstrate a clear case where this correlation cannot hold, for example, as these sources lack the \ion{H}{1} emission that serves as a proxy for accretion rate for the YSOs in our sample \citep{CG10}. The FU Ori phenomenon is sufficiently distinct from the ``standard'' magnetospheric accretion mechanism thought to operate in our Class~II sources, however, that the absence of this correlation in the FU Ori phase is not dispositive of the implied connection between the {\pab} emission, redshifted \ion{He} {1}~$\lambda10830$ absorption, and the underlying accretion mechanism.

The fact that the blueshifted component of \ion{He}{1} is not significantly correlated with accretion activity over time seems, at first blush, to argue against a model wherein YSO winds are instantaneously accretion-powered. However, despite the lack of a formal statistical correlation between the {\pab} and blueshifted \ion{He}{1} EWs, visual inspection of Figure~\ref{fig:hecorr} does reveal a qualitative (although modest) connection between these quantities. Additionally, when considering the three YSOs as a sample, the time-averaged strength of the \ion{He} {1}~$\lambda10830$ blueshifted absorption feature is correlated with the time-averaged accretion rate as inferred from {\pab} and {\brg} emission, in the sense that GY~292 has both the highest {\Macc} and the largest time-averaged blueshifted \ion{He}{1} EW, while GY~314 has the lowest in both. Furthermore, we do find a statistically significant correlation between the total \ion{He}{1}~$\lambda10830$ line strength, defined as $|$EW$_{\mathrm{red}}|+|$EW$_{\mathrm{blue}}|$, and the {\pab} EW in these YSOs, deriving $r=-0.69$ and $r=-0.65$ for GY~292 and GY~308, respectively (we again excluded GY~314 because it lacks {\pab} emission at several epochs). This result agrees with the trend found by \cite{Edwards06}, who noted an increase in both total \ion{He}{1}~$\lambda10830$ line strength and Pa$\gamma$ emission with increasing $Y$~band veiling (and thus a mutually increasing relation between \ion{He}{1} and Pa$\gamma$\footnote[6]{We measure {\pab} instead of Pa$\gamma$ due to our greatly increased S/N at the wavelength of the former as compared to the latter; the relative strengths of these two $n_{\mathrm{lower}}=3$ transitions appears to be consistent in our data (see \S \ref{sec:EmLineVar}).}). Interestingly, the slope of the relation between blueshifted \ion{He} {1}~$\lambda10830$ absorption and {\pab} emission reverses sign (as indicated by the sign of $r$) for GY~292 as compared to GY~308 and GY~314, which suggests further complications in the dynamics of the accretion/outflow relationship. This difference in sign may represent further differences in accretion or outflow geometry between these YSOs, or since the correlations are not statistically significant, the reversal in sign may simply be due to statistical noise.

Detailed time-series studies of more objects, and at higher spectral and temporal resolution, are needed to more fully explore the apparently complicated relationship between stellar outflows and accretion activity. Direct, simultaneous comparisons of the \ion{He} {1}~$\lambda10830$ profile with other accretion and outflow indicators, such as UV-continuum excess, could also help clarify the relative contributions of accretion and outflow to this complex line profile.

\section{Summary}

\begin{enumerate}
\item{We derived spectral types, extinctions, veilings, and mass accretion rates for a small sample of YSOs in $\rho$~Oph. All three are Class II YSOs (CTTSs) and active accretors with time-averaged mass accretion rates of $\sim 10^{-9.2}-10^{-8.5}$~{\Msun}/yr.} Our results suggest an evolutionary sequence (which may not directly map to an age sequence) from least to most evolved of GY~292, GY~308, then GY~314.
\item{IRAC mid-IR light curves were also obtained for these objects as well as the Class~I YSOs YLW~15a and YLW~16b. Three of five of these YSOs exhibit statistically significant variability with [3.6] amplitudes of $\gtrsim 0.2$~mag. Interestingly, GY~292, the Class~II YSO that is the least evolved (i.e. the strongest accretor and most veiled) is the least variable in the mid-IR over day- to week- timescales, suggesting that mid-IR variability amplitude is not monotonic as a function of accretion activity or dominance of the circumstellar disk.}
\item{To investigate possible mechanisms connected with mid-IR variability, we examined the extinction, veiling, and mass accretion rate of each YSO as a function of time. Accretion rates, as traced by \ion{H}{1} emission, are seen to vary by as much as a factor of several over our two month baseline, while variations in the emission by the inner disk, as traced by NIR veiling, are much more modest. However, we cannot directly link changes in either accretion or disk emission to variations in the IRAC light curves. Extinction is relatively constant over our observational baseline, and we can rule it out as the sole cause of mid-IR variability in at least one object, GY~314. Self-obscuration by the disk remains a possible scenario in the other two YSOs; this possibility could be further investigated through higher precision mid-IR photometry and/or polarimetry to detect scattered light}
\item{We find that redshifted \ion{He}{1} $\lambda 10830$ emission, where present in our spectra, shows both quantitative and qualitative similarities to the accretion-sensitive \ion{H}{1} emission lines, suggesting that the winds presumably traced by the \ion{He}{1} line may be connected with accretion activity. Blueshifted \ion{He}{1} absorption, while not formally correlated with {\pab}, is strongest (in a time-averaged sense) in GY~292, the least evolved source, and practically absent in GY~314, the most evolved source. Additionally, the total \ion{He}{1}~$\lambda 10830$ line strength is correlated with the {\pab} EW, further supporting a connection (albeit a complicated one) between accretion, veiling, and winds in YSOs.}
\end{enumerate}

\acknowledgements The authors would like to thank W. Fischer, R. Kurosawa, S. Offner, A. Cody, and E. Chiang for useful conversations, K. Luhman for kindly sharing his spectral data, as well as the anonymous referee, whose valuable comments helped improve this manuscript. We also owe a debt of gratitude to Bill Golish, who helped us obtain our SpeX observations with remarkable efficiency. K.R.C. acknowledges support for this work from the Hubble Fellowship Program, provided by NASA through Hubble Fellowship grant HST-HF-51253.01-A awarded by the STScI, which is operated by the AURA, Inc., for NASA, under contract NAS 5-26555. C.M.F. acknowledges support from the National Science Foundation Graduate Research Fellowship under Grant No. DGE-1144152. 

This work is based in part on observations made with the \textit{Spitzer} Space Telescope, which is operated by the Jet Propulsion Laboratory, California Institute of Technology under a contract with NASA.

This research has made use of NASA's Astrophysics Data System Bibliographic Services, the SIMBAD database, operated at CDS, Strasbourg, France, the NASA/IPAC Extragalactic Database, operated by the Jet Propulsion Laboratory, California Institute of Technology, under contract with the National Aeronautics and Space Administration, and the VizieR database of astronomical catalogs \citep{OBM00}. IRAF (Image Reduction and Analysis Facility) is distributed by the National Optical Astronomy Observatories, which are operated by the Association of Universities for Research in Astronomy, Inc., under cooperative agreement with the National Science Foundation.

The Two Micron All Sky Survey was a joint project of the University of Massachusetts and the Infrared Processing and Analysis Center (California Institute of Technology). The University of Massachusetts was responsible for the overall management of the project, the observing facilities and the data acquisition. The Infrared Processing and Analysis Center was responsible for data processing, data distribution and data archiving.

\clearpage
\appendix
\section{Additional Spectra}
\begin{figure}[h]
\includegraphics[width=0.8\linewidth]{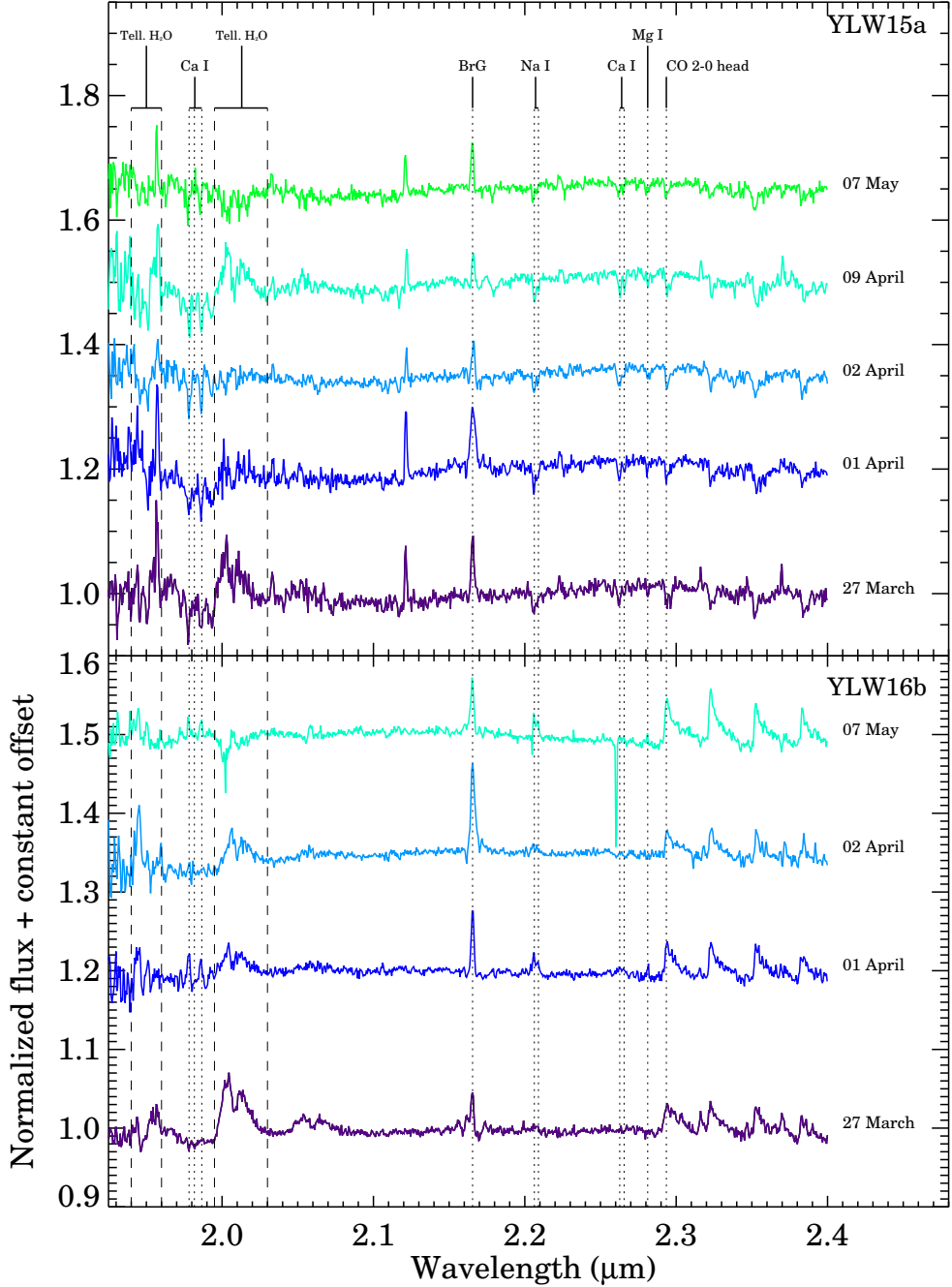}
\caption{$K$~band spectra for the two secondary targets in our sample. Both objects have rising $K$~band slopes, and so the continuum was fit with a second-order polynomial.}
\label{fig:ylws}
\end{figure}

We present the SpeX $K$~band spectra of our secondary targets, YLW~15a and 16b, in Figure~\ref{fig:ylws}.

\end{document}